\documentclass{aa}
\input{psfig.sty}

\begin{document}

\title{Spatially resolved spectroscopy of Cassiopeia A
with MECS \\ on board {\it Beppo}SAX }

\author{
M.C.~Maccarone\inst{1} \and T.~Mineo\inst{1} \and
A.~Preite-Martinez\inst{2} }

\offprints{\\M.C. Maccarone, cettina@ifcai.pa.cnr.it}

\institute{ Istituto di Fisica Cosmica con Applicazioni
all'Informatica, CNR, Via U. La Malfa 153, I-90146, Palermo, Italy
\and Istituto di Astrofisica Spaziale, CNR, Via del Fosso del
Cavaliere 100, I-00133 Roma, Italy}

\date{Received July 13, 2000; accepted December 20, 2000}

\abstract{ We have performed the first detailed spatially resolved
spectroscopy of Cas~A in the 1.6--10 keV energy range, using data
taken with the MECS spectrometer on board the {\it Beppo}SAX
Observatory. The well calibrated point spread function in the
central region of the MECS allowed us to perform a spatial
deconvolution of the data at full energy resolution. We eventually
generated a set of spectra, covering a region of $\sim$3$\arcmin$
radius around the centre of Cas~A. The results obtained by fitting
these spectra using a non-equilibrium ionisation plasma model and
a power law, improve our knowledge about chemical and physical
parameters of the Cas~A SuperNova Remnant: (i) a single thermal
component is sufficient to fit all the spectra; (ii) $kT$ is
rather uniformly distributed  with a minimum in the east and a
maximum in the west, and no evidence is found for high $kT$
expected from the interaction of the main shock with the ISM;
(iii) from the distribution of the values of the ionisation
parameter $n_{\rm e}t$ we infer the presence of two distinct
components: the first ($a$) in the range 1--10 cm$^{-3}$, the
second ($b$) with values ten times higher; if we associate
component $a$ to the CSM and component $b$ to the ejecta, the mass
ratio $M(a)/M(b)\leq$1/10 indicates a progenitor star that lost
only a small fraction of the envelope during its pre--SN life. In
this hypothesis the distribution of component $b$ across the
remnant suggests that the explosion was not spherically symmetric;
(iv) the distribution of abundances indicates that we are
detecting a CSM component with almost solar composition, and an
ejecta component enriched in heavier elements. Abundances found
for $\alpha$--elements are consistent with the current view that
Cas~A was produced by the explosion of a massive star. A low
\element[][]{Fe} overabundance can be an indication that at the
moment of the explosion the mass-cut was rather high, locking most
of the produced \element[][]{Ni}$^{56}$ into the stellar remnant.
}

\titlerunning{Spatially resolved spectroscopy of Cas~A}
\authorrunning{M.C. Maccarone et al.}

\maketitle

\section{Introduction}

\pagenumbering{arabic}

Cassiopeia~A (Cas~A), the youngest known SuperNova Remnant (SNR)
in our galaxy, is an object investigated at several wavelengths.
It has the appearance of a broken shell with average radius of
$\sim$1.7$\arcmin$ (\cite{holt}; \cite{vink96}) around its
expansion center that Reed et al. (1995) found located at
RA$_{2000}$=$23^{\rm h}23^{\rm m}26^{\rm s }.6$,
DEC$_{2000}$=$58\degr49\arcmin01\arcsec$. Detailed observations at
optical wavelengths show a complex morphology, not spherically
symmetric, with two northwest (NW) and southeast (SE) emission
regions moving outward from the remnant center in accordance with
their relative average Doppler motion (\cite{anderson};
\cite{reed}; \cite{lawrence}; \cite{fensen}). A similar
morphological complexity comes out in the X-ray band as observed
by ASCA at the moderate spatial resolution of 3$\arcmin$
(\cite{holt}; \cite{fujimoto}; \cite{vink96}; \cite{hwang97}).
{\it Beppo}SAX (\cite{boella97a}) observed the source giving
interesting results  either in the spectral and in the spatial
analysis. In particular, the  maps obtained  in the energy range
1.6--10 keV (\cite{maccarone98a}; \cite{vink99}) with the Medium
Energy Concentrator Spectrometer MECS instrument
(\cite{boella97b}) show that the spatial distributions of
\element[][]{Si}, \element[][]{S}, \element[][]{Ar}, and
\element[][]{Ca} emission lines do not differ very much each
other.  In addition, the presence of \element[][]{Fe} seems to
extend further east than for the other elements. Such results have
been confirmed recently by the Chandra X-ray observatory that,
with its imaging capability of $\simeq$0.5$\arcsec$, provided well
detailed maps of the remnant and discovered a point-like source
near its center (\cite{hughes}; \cite{chakrabarty};
\cite{hwang00}).

The broad band {\it Beppo}SAX spectrum of the whole source
(\cite{favata}) presents a high energy component well modeled by a
power law of spectral index 2.95. This component has been
localized by Vink et al. (1999) mainly in the southern and western
regions. Indeed the MECS spectra of these regions are somewhat
harder than the average and almost featureless above 4 keV.

In this paper we present results about the spatially resolved
spectroscopy of Cas~A in the MECS energy range 1.6--10 keV, at the
resolution of 1.5$\arcmin$. Data set and data reduction are
presented in Sect.~\ref{Observations-and-Data-Reduction} together
with a brief description of the software environment. Image and
spectral analyses are described in Sect.~\ref{Image-Analysis} and
Sect.~\ref{Spectral-Analysis}, respectively. Results are discussed
in Sect.~\ref{Discussion} while conclusions are presented in
Sect.~\ref{Conclusion}. The treatment of errors is in
Appendix~\ref{Error-Treatment}.

\section{Observations and Data Reduction}
\label{Observations-and-Data-Reduction}

The Medium Energy Concentrator Spectrometer MECS is one of the
narrow field instruments on board {\it Beppo}SAX. The MECS
consists of three units (ME), each composed of a grazing incidence
mirror unit with a position sensitive gas scintillation
proportional counter located at the focal plane. The MECS field of
view covers a radius of 28$\arcmin$ with position resolution of
$\sim$1.5$\arcmin$; inside the central circular window of
$\sim$10$\arcmin$ (delimited by a strongback structure) all the
detector parameters are well defined.

The three ME units were all together active till May 6, 1997, when
ME1 unit was switched off, due to a failure in the high voltage
supply of the gas cell unit. ME2 and ME3 units continued to work
normally.

The on-axis Point Spread Function (PSF) of the MECS can be modeled
as sum of two components: a gaussian, and a generalized
lorentzian. For each ME unit, at a given energy and at a given
distance $r$ from the detector center, the complete analytical
expression for the on-axis PSF is (\cite{boella97b}):

$$ PSF(r,E,ME)~=~{1\over 2\pi \Bigl [ R\sigma^2+{r_l^2\over
2(m-1)} \Bigr ] } ~ \times $$
$$ \Biggl\{ R\exp \Biggl ( -{r^2 \over 2\sigma^2}
           \Biggr ) ~+~ \Biggl [ 1~+~\Biggl ({r \over r_l}\Biggr )^2
           \Biggr ]^{-m} \Biggr\} $$

\noindent where $R$, $\sigma$, $r_l$ and $m$ depend on the ME unit
and are algebraic functions of the energy. The dependence from the
unit can be averaged in the formulation: this allows us to
consider a single PSF at each given energy. Moreover, the on-axis
PSF can be considered valid within a radius of 12$\arcmin$ from
the detector center (\cite{molendi}; \cite{chiappetti}).

Table~\ref{table1} lists the data set reference codes used for the
analysis presented in this paper. MECS data were reduced by using
a standard
procedure\footnote{http://www.sdc.asi.it/software/cookbook} with
appropriate selection criteria so to avoid the South Atlantic
Anomaly and to obtain the maximum rejection of background
particles. A further selection was applied to take into account
only data related to the orbital periods in which the Z star
tracker (aligned with the narrow field instruments) was in use; in
this case the attitude reconstruction achieves an uncertainty of
the order of 30$\arcsec$ inside the strongback region. Under these
conditions, the total exposure time for the MECS was
$\sim$128\,000 s, the net exposure time for each observation is
listed in Table~\ref{table1}.

\begin{table}[h]
\centering \caption{The Cas~A data set used in this paper.}
\label{table1}
\begin{tabular}{cccccc}
\noalign{\smallskip}
\hline
Obs. & Observ. & Observ. & Archival &
Exposure & ME\\
 Prog. & Date & Period & Code & (s) & units\\
\noalign{\smallskip}
\hline
SVP & 06aug96 & 743 & 30011002 & 27574
& 1,2,3\\
SVP & 07aug96 & 745 & 30011001 & 26639 & 1,2,3\\
SVP &
11sep96 & 899 & 30011003 & 16629 & 1,2,3\\
SVP & 11sep96 & 901 &
30011003 & 16595 & 1,2,3\\
 AO1 & 26nov97 & 2990 & 30159001 & 40789
& 2,3\\
\noalign{\smallskip}
\hline
\end{tabular}
\end{table}

Data reduction has been carried out using the {\tt SAXDAS v.1.2.0}
package, developed under the {\tt FTOOLS} environment. Image
analysis and spectra generation were carried out under {\tt
MIDAS-97NOV} environment\footnote{http://www.eso.org/midas}; the
context {\tt
saxmecs}\footnote{http://www.ifcai.pa.cnr.it/$\sim$sax/saxmecs/saxmecs.html}
has been used to convert data in {\tt MIDAS} format, to merge MECS
observations in images, to generate the PSF as function of energy,
and to accumulate the final spectra.

The adopted deconvolution procedure is based on the Lucy's
deconvolution method (\cite{lucy}), as implemented in {\tt
MIDAS-97NOV}, and it can be fruitfully applied on sources
presenting an extended and structured morphology. With respect to
our previous analysis (\cite{vink99}), we reconsidered the
deconvolution parameters: in particular, we rebinned data to
24$\arcsec$ per pixel and performed deconvolution at each energy
channel with the monochromatic PSF computed for that channel, as
detailed in Sect.~\ref{Image-Analysis}.

The spectral analysis has been performed with the
{\tt XSPEC v.9.00} software package.

\section{Image Analysis}
\label{Image-Analysis}

To enhance the morphology of the Cas~A spatial emission in the
energy range 1.6--10 keV at the resolution of the MECS Pulse
Invariant (PI) channel, we adopted the following procedure: for
each PI channel in the chosen energy range (PI channel from 32 to
214) we have (i) selected data from each observation, (ii)
converted data in X,Y sky pixels images, (iii) merged and rebinned
images to improve statistics, (iv) generated the appropriate PSF.
We then deconvolved each monochromatic rebinned image with its
proper monochromatic PSF for a suitable number of iterations.

Rather large inaccuracies are present in the aspect solution of
our original Cas~A data set, particularly in the last observation
(AO1) made after a gyroscope failure. To overcome the problem,
present whenever we need to merge spatial data coming from
different observations, we used a procedure similar to that
adopted in Vink et al. (1999) and detailed in Maccarone \& Mineo
(1999)\footnote{
http://www.ifcai.pa.cnr.it/$\sim$cettina/papers/CasA-report2.ps.gz}.

\subsection{The Deconvolution Algorithm}

To be successfully applied, the Lucy's deconvolution algorithm
needs to be properly tailored taking into account the
PSF characteristics, the spatial region of Cas~A we are
interested in, the statistical significance level we are dealing with,
and the number of iterations necessary to reach
convergence.

The spatial region of Cas~A we are interested in is all inside the
inner strongback window and it is centered very near to the
detector center. Therefore, the radial description of the on-axis
PSF can be successfully used in the deconvolution process. We
applied the deconvolution algorithm to each PI channel within the
chosen energy range by using the monochromatic PSF corresponding
to that channel. The total number of counts within the interested
region is about $3.7\times10^6$ distributed over all the PI
channels with a maximum of $\sim1.5\times10^5$ and a minimum of
$\sim300$ counts.

To improve statistics, we compacted both PSF and observational
data images by a factor of 3. This corresponds to rebin their
stepsize from 8$\arcsec$/pixel to 24$\arcsec$/pixel, and to obtain
a well-deconvolved spatial region of $\sim$10$\arcmin$ radius.

To define what number of iterations must be used in the
deconvolution algorithm, we adopted a convergence criterium based
on the relative change between images produced by two consecutive
iterations. The iterations can be stopped when a given level of
convergence for the maximum relative change is reached. From
previous checks (\cite{maccarone98b}; \cite{vink99}), we noted
that a maximum of 30 iterations is sufficient to enhance the
morphology of the Cas~A emission regions at higher energies
(further iterations produce small corrections that tend to match
the statistical fluctuations) while, at lower energies a greater
number of iterations (some hundreds) is necessary to reach
convergency.  In the case of the Cas~A data images analyzed here
with stepsize of 24$\arcsec$/pixel at the resolution of the MECS
PI channel, a number of 200 iterations is sufficient to reach
convergency at lower energies. To automatize the full
deconvolution procedure we then choose to apply 200 iterations for
any energy channel, without any risk of compromising the results
for the higher energy channels. This choice was also supported by
the comparison (evaluated via a standard $\chi^2(i)$ function at
each {\it i}-th iteration) between the input image and the
convolution of the {\it i}-th deconvolved image with its proper
PSF (Maccarone \& Mineo, 1999).

\begin{figure}
\centerline{ \hbox{ \psfig{figure=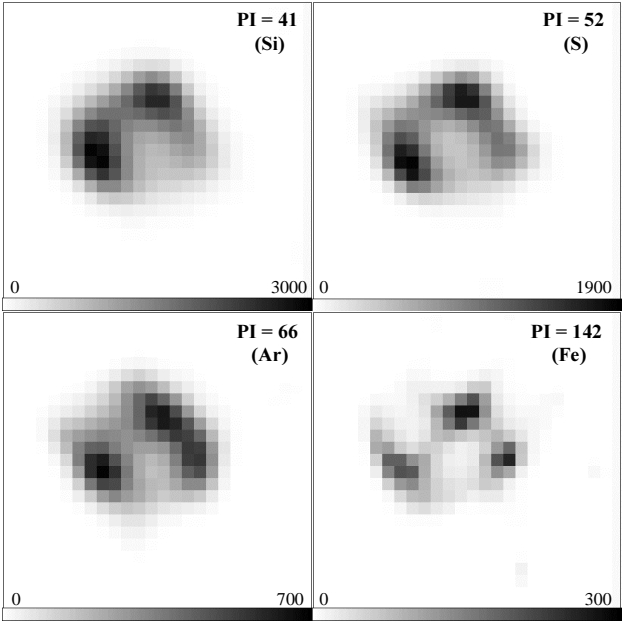,width=8.5cm,clip=} }}
\caption{Deconvolution of the images related to PI channels 41,
52, 66, and 142, corresponding to $\sim$ 1.89, 2.41, 3.06, and
6.66 keV, respectively.} \label{fig1}
\end{figure}

Fig.~\ref{fig1} shows the results after 200 iterations on the
images at the energy of \element[][]{Si}, \element[][]{S},
\element[][]{Ar}, and \element[][]{Fe} lines. The four
monochromatic images present the same morphological structure
already illustrated in Fig. 3, left panels of our previous paper
(\cite{vink99}), although the setting of the deconvolution
parameters are sligthly different: 3$\times$3 compacted pixels and
full spectral resolution in this paper, 2$\times$2 compacted
pixels and wide energy bands in our previous paper. From the image
analysis point of view, the setting used here gives results in
perfect agreement with those obtained in our previous paper;  at
the same time this setting allows us to perform a better spatially
resolved spectroscopy at $\sim$1.5$\arcmin$ resolution level.

\section{Spectral Analysis}
\label{Spectral-Analysis}

\begin{figure}
\centerline{ \hbox{ \psfig{figure=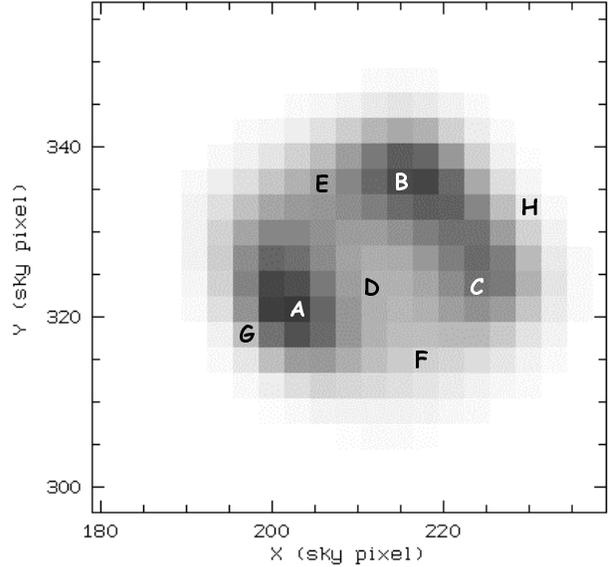,width=8.5cm,clip=} }}
\caption{X,Y projection of the Cas~A data cube. The total counts
range from 800 to $\sim$62000 counts/pixel. Labels refer to
spectra shown in Fig.~\ref{fig3}.} \label{fig2}
\end{figure}

The set of Cas~A images obtained from the deconvolution procedure
has been organized in a position-energy cube (X, Y, E) from which
we have accumulated spectra relative to each X,Y pixel in the
spatial region of interest. To work with a good statistical signal
we have only considered spectra with total counts greater than
800, reducing to 182 the final number of spatially resolved
spectra to be analyzed. The window on which we have performed the
spatially resolved spectroscopy then covers a region of
$\sim$3$\arcmin$ radius around the Cas~A expansion center. In sky
pixels, the coordinates of this center are X$=$209, Y$=$327.
Fig.~\ref{fig2} shows the map of the total counts of the 182
selected pixels.

Statistical errors on the counts have been increased to take into
account the systematics of the deconvolution procedure (see
Appendix~\ref{Error-Treatment}); spectra have been rebinned to
have at least 20 counts per energy bin.

Cas~A is a strong source and the MECS background subtraction is
not a relevant problem even if the deconvolution procedure acts on
the spatial distribution of the background, moving counts from the
external regions toward the center. We evaluated the background
level in the worst case for which all counts within the
well-deconvolved region (10$\arcmin$) are placed under the source.
Under the simplified hypothesis of an uniform background, the
level is of the order of $10^{-4}$ counts/s/pixel. We then ignored
the background subtraction in our spectral analysis.

In the fitting procedure we used the MECS response matrix relative
to an on-axis point source with an extraction circle of
10$\arcmin$ radius for all spectra, while the deconvolution was
applied using a PSF defined in a squared area: this difference in
geometry introduces an under-estimation of the effective area less
than 1\% in the spectral analysis.

\begin{figure*}[t]
\centerline{ \hbox{ \psfig{figure=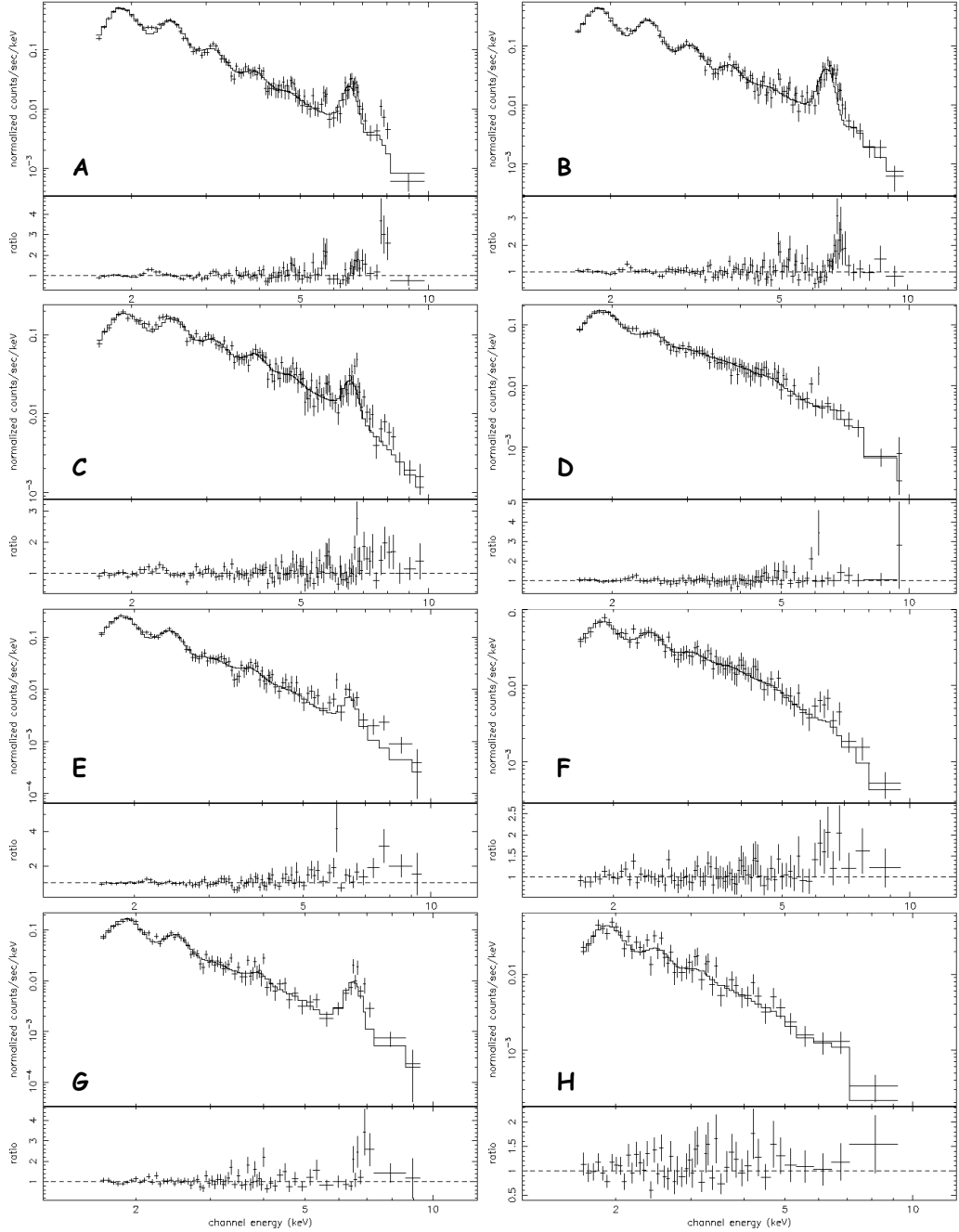,height=18cm,clip=} }}
\caption{Spatially resolved spectra of the pixels marked with a
letter in Fig.~\ref{fig2}. For each frame, top panel refers to
data and folded model, while bottom panel refers to the fitting
residuals in unit of data-to-folded model ratio.} \label{fig3}
\end{figure*}

Fig.~\ref{fig3} shows a selection of spectra extracted at the
positions marked with the corresponding letter in Fig.~\ref{fig2}.
It is evident that Cas~A emission has not uniform spectral
characteristics. In particular, as suggested by Vink et al. (1999)
and Hughes et al. (2000), two main emission mechanisms can be
identified: a thermal process, underlined by the presence of
emission lines and coincident with the broken shell
(Fig.~\ref{fig3}-A, B, C, E, G), and an emission process
characterized by the absence of strong lines (Fig.~\ref{fig3}-D,
F, H) and essentially located in the central region between the
main peaks (Fig.~\ref{fig2}).

To fit all spectra, we properly implemented in the fitting code a
model of emission from a non-equilibrium ionization (NEI) plasma.
The low energy absorption $N_{\rm H}$ has been constrained within
the range 0.9--1.7$\times$10$^{22}$ cm$^{-2}$, slightly wider than
the range found by Keohane et al. (1996). The abundances of
\element[][]{Si}, \element[][]{S},
\element[][]{Ar},\element[][]{Ca},\element[][]{Fe} and
\element[][]{Ni} have been left as free parameters and
 allowed to vary in the range 0.02--20 (with respect to solar).
Moreover we added to the thermal model a power law to take into
account the 20--80 keV emission found by Favata et al. (1997). The
parameters of the power law are not free in the fitting procedure.
In fact, none of the MECS deconvolved spectra has enough
statistical significance to justify the presence of a second
spectral component (e.g., the power law) superimposed to the
thermal emission model.

The spectral index has been fixed at the value of 2.95 as quoted
in Favata et al. (1997); the normalization  has been computed by
assuming that the whole remnant contributes to the high energy
component and that the emissivity of this component is
proportional to the MECS flux detected in each pixel. In the
framework of the present analysis, we realized that the hard tail
cannot be located in the two regions, south and west, suggested in
our previous analysis (\cite{vink99}). In fact, by fitting with a
single power law in the 4--10 keV range the continuum emission of
the spectra of these regions (covering $\simeq50\%$ of the
remnant), and extrapolating the power laws in the high energy
range, we found that the extrapolated flux is insufficient by a
factor $\sim$2 to justify the high energy detection.

While the paper was in the refereeing process, results from
XMM-Newton observations of Cas A became available
(\cite{bleeker}), fully confirming our assumption. Indeed they
states: "the hard x-ray image and the hardness ratio indicate
that" the 8.1--15 keV "flux does not predominate in a few
localized regions, but pervades the whole remnant in a
distribution similar to the softer thermal component".

Most of the spectra  are well fitted by the composite model: the
$\chi^2$ values are within the expected range and no systematic
residuals are observed in any of the spectra (see
Fig.~\ref{fig3}).

\section{Results and Discussion}
\label{Discussion}

\begin{figure*}
\centerline{ \hbox{ \psfig{figure=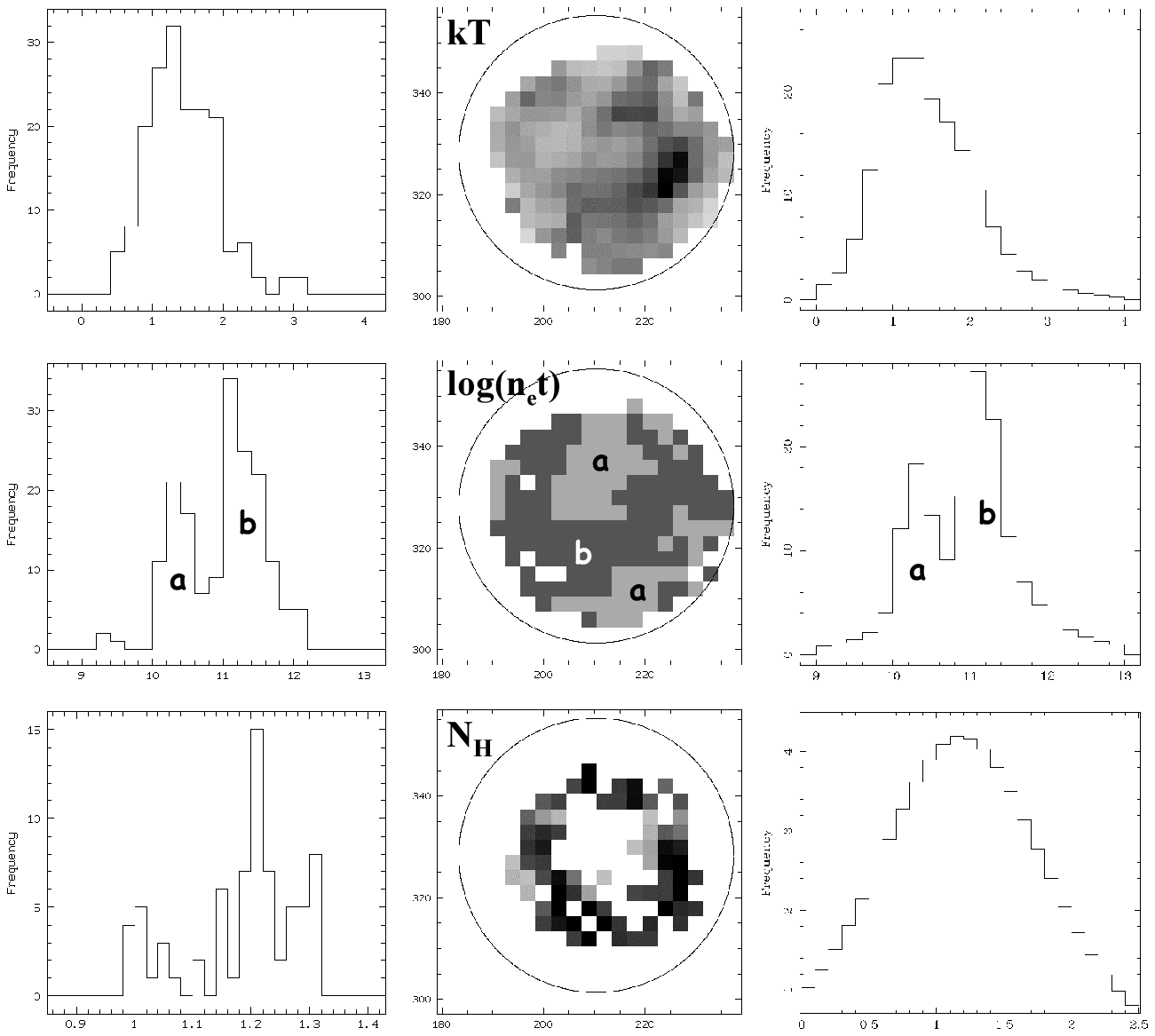,width=13cm,clip=} }}
\caption{NEI plasma model: frequency and spatial distribution of
$kT$, $\log(n_{\rm e}t)$, and $N_{\rm H}$. The superimposed circle
is only for positional reference. The right panels show the
frequency distribution of the same data with errors taken into
account (see text).} \label{fig4}
\end{figure*}

Results obtained fitting the 182 spectra with a NEI plasma model
are listed in Table~\ref{table2}. They are also shown in
Fig.~\ref{fig4} and in Fig.~\ref{fig5}, where we present  an
histogram of the frequency distribution (left panels) and a map of
the spatial distribution of the best fit values for each variable.
In addition, we present in the right panel of Fig.~\ref{fig4} and
Fig.~\ref{fig5} a frequency histogram of the same data used in the
left panel, but where each single determination is transformed
into a normalized gaussian with central value and $\sigma$ as
derived from the fits.

We note that in previous studies analyzing data from a variety of
different instruments including BeppoSAX (\cite{hill}, from a
rocket-borne proportional counter; \cite{davison}, experiment C on
{\it Ariel--5}; \cite{pravdo}, proportional counter on OSO--8;
\cite{becker}, SSS on {\it Einstein}; \cite{jansen}, EXOSAT;
\cite{holt}, ASCA; \cite{vink96}, ASCA; \cite{favata}, {\it
Beppo}SAX), reasonable fits of the spatially unresolved spectrum
of Cas~A were obtained using two thermal models with different
physical and chemical parameters, interpreted as an indication of
the presence of two different components in the remnant. The only
exception is given by the analysis of {\it Tenma} data (Tsunemi et
al. 1986). In particular, BeppoSAX unresolved spectrum has been
modeled with two NEI components respectively at the temperature of
1.25 and 3.8 keV. Our spatially resolved spectroscopic analysis
allowed us to get statistically significant fits using only one
thermal component per pixel with temperatures compatible with the
one relative to the cooler component in Favata et al. (1997). This
means that if two thermal components are indeed present, they are
spatially separated.

\subsection{$kT$}

$kT$ values are distributed in the range 0.9--3.2 keV, with most
of the values in the rather narrow 1--2 keV range. The
distribution broadens slightly if we take into account the
uncertainties on the derived values (Fig.~\ref{fig4}, right
panel). The shape of the continuum does not vary considerably
across the remnant. We notice three exceptions: (i) a well defined
minimum NE of centre with $kT\sim$0.9 keV, (ii) a maximum W of
centre with $kT\sim$3 keV, correlated with the relative maximum in
total emission indicated as "C" in Fig.~\ref{fig2}, and (iii) low
values at the edge of the remnant (with the exception of the
southern edge). A hot western component was also found by Vink et
al. (1996) analyzing their ASCA data of Cas~A.

It is well known from the physics of strong shocks through a gas
with adiabatic index $\gamma=$5/3, that electron temperatures in
the range 1--2 keV can be reached by post-shocked matter when the
shock velocity is of the order of 900--1300 km s$^{-1}$. From
observations in the optical (\cite{reed}) and in the X-rays
(\cite{holt}; \cite{vink98}) we can infer that the main shock of
Cas~A is presumably running now  more than 2 times faster,
producing post-shock temperatures over 4 times higher than
observed. Even in the hypothesis of a recent deceleration due to
the interaction with a denser ISM, we should see in the $kT$ map
the signature of such a deceleration in the form of a substancial
increase of $kT$ at the edge(s) of the remnant. On the other hand,
higher resolution X-ray images clearly show a strong surface
brightness contrast between the main shock and the bright ring
that dominates the X-ray and radio emission. Nowhere in our 182
locations across the remnant we detect  high temperatures,
strongly indicating that the main shock is now interacting with a
very tenuous ISM, generating an emission that is too faint to be
detected in our observations.

Although the morphology of the distribution of $kT$ values is
certainly of some importance, the real question to ask is: why do
we detect a $kT$ in the range 1--2 keV? A correlated question is:
why do we see matter enriched in heavy elements (\element[][]{Si}
through \element[][]{Ni}) emitting in the X-rays at such
temperatures? A similar question will be asked further on in a
different context and for different reasons. What we know about
the hydrodynamics of the explosion inside a star is rather well
established: the shells of the exploding star closer to the
mass-cut (e.g. the mass coordinate separating what will be ejected
from the stellar remnant), carrying out the yield of explosive
nucleosynthesis, reach extremely high temperatures fractions of a
second after the beginning of the explosion inside the star. Few
hours later, the main shock is still inside the star, at the base
of the photosphere, ready to break through giving rise to the
standard SuperNova event in the V and UV bands. At the same
moment, those inner shells pulled out by the main shock have
expanded almost to the surface of the star. But the key point is
that they have already cooled down to temperatures lower than 0.01
keV.

If emission is dominated by the ejecta (but \cite{borkowski}
suggest a different interpretation), how to re-heat those ejected
shells? The simplest answer can be found in Astrophysics textbooks
and reviews (e.g.: \cite{mckee}): ejected matter acts like a
piston, a secondary shock forms, running back (in lagrangian
coordinates) into the ejecta, re-heating it to temperatures lower
than those produced by the main forward shock. We just note {\it
en passant} that nobody thought of shifting the onset of secondary
shocks to places where the probability of formation is certainly
high (inside the star), and to the right time (hours before the
SuperNova event).

Being this not the right place to question the applicability of this
model to this particular case, we limit ourselves to suggest a
different, more straightforward scenario, based on what we observe
also in other wavelength ranges. The basic ingredients of our
scenario are azimuthal secondary shocks to convert a small
fraction of the bulk kinetic energy ($\sim10^{51}$erg) of
"polluted" shells into thermal energy. Only few \% of kinetic
energy need to be converted by azimuthal shocks into thermal
energy to set up the scene for what we will observe 10$^{10}$s
later. Meanwhile, vorticity will help forming blobs or knots of
frozen local composition and scrambling their relative positions.

The scenario is the following: at time $t \simeq 0$ (plus or minus
hours, does not really matter) we deal with shells of material
enriched in heavier elements that are rather cold. Energy is
mostly in the form of kinetic energy. Their bulk motion is
subsonic with respect to adjacent shells, but supersonic with
respect to the value of the sound speed in the shells. This is a
very unstable situation. Any element of matter with a random
azimuthal velocity component will be shocked (or compressed) by radially
moving adjacent matter. Kinetic energy will be transformed into thermal
energy, and vorticity will set in. Of course the complexity of
this basic scenario,  based on azimuthal shocks in a cold
supersonic flow, can be increased {\it ad libitum} adding
rotation, and in general any non spherically symmetric effect.

\subsection{$N_{\rm H}$}

Due to the limited extension towards low energies of the MECS well
calibrated band (E$\geq$1.6 keV) we cannot determine $N_{\rm H}$
with reasonable accuracy. Taking into account the uncertainties in
the determination of $N_{\rm H}$ (see Fig.~\ref{fig4}, right
panel) we can say that our results are consistent with those of
Kehoane et al. (1996). Moreover, we fitted one of the most
significant spectra (namely, the spectrum marked with "A" in
Fig.~\ref{fig3}, corresponding to the pixel X=203, Y=321 in
Table~\ref{table2}) fixing $N_{\rm H}$  to three different values;
the results, as listed in  Table~\ref{table3}, show that no
significant variations in any of the  $N_{\rm H}$ sensitive
spectral parameters are detected.

\begin{table}[h]
\centering \caption{Test for the spectrum marked with "A" in
Fig.~\ref{fig3} of the sensitivity of the derived parameters to
$N_{\rm H}$ . Quoted errors are one standard deviation.}
\label{table3}
\begin{tabular}{llll}
\noalign{\smallskip} \hline

~~~$N_{\rm H}$ (10$^{22}$ cm$^{-2}$) & \multicolumn{1}{c}{0.7}
&\multicolumn{1}{c}{1.3}
 &\multicolumn{1}{c}{1.9} \\ \noalign{\smallskip} \hline

~~~kT (keV) & 1.6$\pm$0.26 & 1.46$\pm$0.29 &1.4$\pm$0.22 \\

~~~log($n_{\rm e}t$)(cm$^{-3}$s) & 11.25$\pm$0.2  & 11.3$\pm$0.22
& 11.3$\pm$0.2 \\

~~~[\element[][]{Si}] & 5.9$\pm$1.35 & 6.46$\pm$1.72 &8$\pm$2
\\

\noalign{\smallskip} \hline
\end{tabular}
\end{table}

\subsection{$n_{\rm e}t$}

The ionisation parameter $n_{\rm e}t$ controls the non equilibrium
ionisation status of the emitting plasma. Because we express
$n_{\rm e}t$ in units of cm$^{-3}$s and its value can vary by
orders of magnitude, in Table~\ref{table2}, in Fig.~\ref{fig4} and
in this discussion we will always refer to values of $\log(n_{\rm
e}t)$. For the elements considered in this paper (\element[][]{Si}
through \element[][]{Ni}) a low value of the ionisation parameter
(e.g. $\log(n_{\rm e}t)<10.5$) indicates an ionisation structure
far from equilibrium, while the structure is practically in
equilibrium for values greater than 12. In the fitting procedure
$\log(n_{\rm e}t)$ is allowed to vary in the range 6--12.5.

The frequency distribution of $\log(n_{\rm e}t)$ values is clearly
bimodal, with lower boundaries at about 10 and 11. Because an
upper limit for $t$ is the age of the remnant (assumed to be 320
y, e.g. $\log(t)$=10 in our units), this can give us a good handle
in understanding the density distribution within the remnant. The
simplest hypotheses concerning the density distribution are:

case (i): we fix $t$ to infer the distribution of $n_{\rm e}$. We
are observing matter shocked (once or more than once) 10$^{10}$s
ago, that is during the early phases of the explosion. Then the
distribution of the ionisation parameter actually reflects the
distribution of the electron density, shifted by 10 orders of
magnitude. If this hypothesis is true, we detected a bimodal
density distribution, with a first component (marked $a$ in
Fig.~\ref{fig4}) with $n_{\rm e}$ in the range 1--10 cm$^{-3}$ and
a second component (marked $b$ in the same Figure) about 10 times
denser. All in all $n_{\rm e}$ (actually: the electron density
averaged over our pixel volume) would cover the range 1--$>$100.

case (ii): we fix the distribution of $n_{\rm e}$ to infer the
distribution of $t$ (e.g. of the main shocking episodes). The
sharp boundary at $\log(n_{\rm e}t)$=11 indicates that in the
remnant is present only one component ($b$) with (pixel averaged)
$n_{\rm e}$ in the range 10--100 cm$^{-3}$ and over that was
shocked 10$^{10}$s ago. From the boundary at $\log(n_{\rm e}t)$ =
10 we can infer that a consistent fraction of this same matter
(with $n_{\rm e} \geq$ 10 cm$^{-3}$) was suddenly re-shocked
10$^{10}$ cm$^{-3}s$/10 cm$^{-3}$ = 10$^9$s $\simeq$ 32 y ago
(that is in the mid sixties, when Cas~A was already well
observed). There could have been two different causes for this
sudden re-shocking of part of the remnant: external, e.g. matter
at the edge of the remnant hit pre-existing clouds or
condensation; internal, e.g. blobs or fingers of material inside
the remnant collided and re-heated. What we observe in
Fig.~\ref{fig4} is that this hypothetically re-shocked material is
located mostly at the centre of the remnant and north of it, with
few other pixels at the edges. But no particular anomalies in $kT$
are found at the corresponding locations that could be interpreted
as signatures of a very recent re-shocking.

These first two cases are the extreme possibilities in which we
fix either $t$ or $n_{\rm e}$.

(iii) a third intermediate possibility is that matter is
distributed in components $a$ and $b$ as in (i), plus an
additional component $c$ of density higher than component $b$ by a
factor $n$, that was then re-heated  $age/n$ years ago. Factor $n$
must be significantly greater than 1 in order to consider this
case different from case (i).

The above hypotheses have an obvious impact on the determination
of the total mass $M$ of the remnant: $M(ii)$ can be up to 4 times
$M(i)$, and $M(i)<M(iii)<M(ii)$.

We favour hypothesis (i) on the basis of the following
considerations:

(1) case (i) is not in contrast with the other results of our
analysis. On the other hand we don't see evidence for the
re-heating implied by cases (ii) and (iii).

(2) the observed values of $kT$ are much too low to indicate the
presence of shocked ISM (i.e. matter not ejected by the progenitor
star in its pre-SN life). These values imply that what we are
observing is just matter belonging to the progenitor star spread
out in a volume of about 3 pc in radius.

(3) the lowest possible total mass of the (X-ray) emitting gas
computed according to hypothesis (i) is already in the range
13--30 M$_\odot$, reasonable masses for the progenitor star. By
the way, X-ray emitting masses in this range were already
suggested by Murray et al. (1979) and Fabian et al. (1980).

Combining the two last considerations, one sees that no much room
is left for hypothesis like (ii) or (iii), that will lead to total
masses exceedingly large for the progenitor star.

If we then retain only hypothesis (i), component $a$ could be
associated to the CSM, and the denser component $b$ to the
disrupted star. We should then detect a correlation between the
spatial distribution of component $b$ and that of overabundances.
Anticipating the results of the discussion of Fig.~\ref{fig5}, we
can say that this hypothesis is not in contradiction with the maps
of abundance values shown in Fig.~\ref{fig5}, with the only
exception of the northern sector of the remnant.

Another interesting consequence concerns the ratio by mass of
these two components. Although components $a$ and $b$ are spread
over $\sim$40\% and $\sim$60\% of the pixels, respectively, the
mass ratio $M(a)/M(b)= M(CSM)/M(ejecta)$ is $\leq1/10$. This ratio
is much smaller than in previous estimates (see \cite{vink96},
\cite{borkowski}). In other words, our results seem to suggest
that the progenitor star lost in its pre-SN phase a small fraction
of its mass. If Cas~A was an underluminous SN event, it wasn't
because it lost most of its envelope before the explosion. Other
explanations should be found: a moderate initial mass or a high
mass-cut at the moment of the explosion, if component $a$
represents the total CSM mass, or, alternatively, we are detecting
only a fraction of the total CSM, which is actually distributed
over a much larger volume than the remnant. This would be the case
if much of the mass loss occurred when the progenitor was a blue
supergiant.

We also remark that the denser component is not distributed in our
map (in Fig.~\ref{fig4}) as it should if matter had been ejected
in spherical symmetry. On the contrary, the shape of the
distribution of component $b$ in Fig.~\ref{fig4} could support the
hypothesis of a toroidal ejection with an axis of symmetry
oriented from NE to SW. Markert et al. (1983) were the first to
notice from X-ray spectra of Cas~A, taken with the crystal
spectrometer on board the {\it Einstein} Observatory, that the
northwest part of Cas~A is red-shifted with respect to the
southeast, with a dividing line running then from NE to SW. The
same symmetry was found by Holt et al. (1994) in their ASCA data,
from which they could build a Doppler contour map of Cas~A showing
that the SE part of the remnant was blue-shifted and the NW region
was red-shifted. The torus could be by now broken in a number of
different concentrations, giving rise to the SE emission
enhancement (running towards us) and other in the north and west
(running away from us).

\subsection{Abundances}

Abundances were derived, with varying degree of accuracy, for
\element[][]{Si} in 167 out of the 182 pixels, for \element[][]{S}
in 164 pixels, for \element[][]{Ar} in 106 pixels, for
\element[][]{Ca} in 90 pixels, for \element[][]{Fe} in 105 pixels,
and for \element[][]{Ni} in 49 pixels. The determination of
\element[][]{Si} and \element[][]{Ni} abundances is affected by
the position of the corresponding blends at the opposite extremes
of our energy window. The fit of the \element[][]{Si} blend is
affected by the unability to constrain reasonably well the value
of $N_{\rm H}$, while the \element[][]{Ni} blend is located in a
region of the spectrum where the count rate is the lowest,
although the presence of a \element[][]{Ni} line is not related to
the statistics of the spectrum: see spectra A and B in
Fig.~\ref{fig3}. The list  of elements in order of decreasing
global accuracy of the corresponding abundance determination is:
\element[][]{S}, \element[][]{Fe}, \element[][]{Si},
\element[][]{Ar}, \element[][]{Ca}, and \element[][]{Ni}.

\begin{figure*}
\centerline{ \hbox{ \psfig{figure=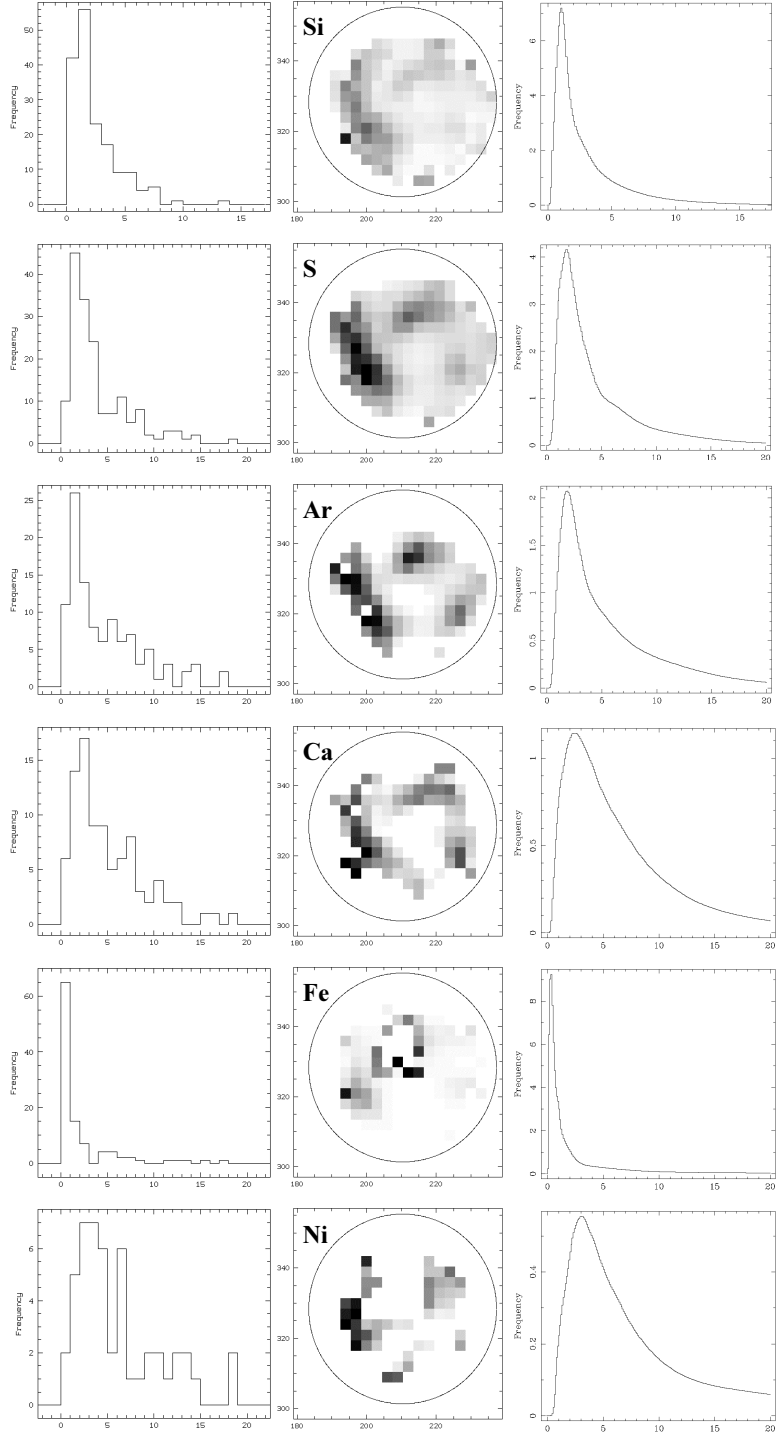,height=19.5cm,clip=}
}} \caption{NEI plasma model: frequency and spatial distribution
of \element[][]{Si}, \element[][]{S}, \element[][]{Ar},
\element[][]{Ca}, \element[][]{Fe}, and \element[][]{Ni}
overabundances. The superimposed circle is only for positional
reference. The right panels show the frequency distribution of the
same data with errors taken into account.} \label{fig5}
\end{figure*}

The frequency distributions of abundance determination (with
respect to solar, by number) are shown in the left panels of
Fig.~\ref{fig5}. The right panels map the spatial distribution of
these values, with the same intensity scale for all six panels.
The same reference circle centered at the expansion centre (as
defined by \cite{reed}) is shown in all panels for sake of
comparison of the different distributions.

Now, we know that Cas~A is located not much further away from the
centre of the Galaxy than our Sun, and was generated by the
explosion of a massive star born between about 3 to 10 My ago. In
terms of the time scale of galactic evolution, this means "now",
and today's abundances of the ISM at the location of Cas~A are
basically solar. On all this there is a general consensus. As well
as on the obvious expectation to detect overabundances of heavy
elements (in our case: \element[][]{Si} to \element[][]{Ni}) with
respect to solar values, as found in the optical knots.

But from common knowledge (certainly still limited) of explosive
nucleosynthesis the following question is fully justified: should
we really expect to observe those overabundances? The answer is:
yes, for $\alpha$-elements, and: not necessarily, for
\element[][]{Fe} and \element[][]{Ni}. Suppose the extreme
hypothesis of a complete mixing, above the mass-cut, of all the
matter belonging to the progenitor star. The production of
$\alpha$-elements is such that, unless stellar matter is diluted
in a much greater mass of ISM, we do expect to observe
overabundances of \element[][]{Si} through \element[][]{Ca}. On
the other hand, depending on the precise value of the mass-cut,
overabundances of \element[][]{Fe} and \element[][]{Ni} could well
pass undetected.

A more likely scenario, based on high spatial resolution studies
of this remnant in the optical (and now also in the X-rays), is
that Cas~A is a remnant composed by scrambled "blobs" or knots of
frozen local material more than a mixed distribution of matter. In
this case we should re-formulate our answer to the previous
question, saying that: yes, we do expect overabundances of heavy
elements, but only in the location of parcels or blobs of ejecta
keeping trace of the effects of explosive nucleosynthesis. In
other locations we can only expect to find abundances that will
reflect those of the ISM at the time the progenitor star was born.
Actually, due to s-process nucleosynthesis, the \element[][]{Fe}
abundance could even be depleted below its initial value (see for
instance the recent paper by \cite{limongi}).

From the left panels of Fig.~\ref{fig5} we clearly see that in the
majority of the pixels we are detecting abundance values
compatible with solar values. More in detail, there seems to be
indications that in the location of Cas~A abundances of
\element[][]{Fe} and \element[][]{Ni} are somewhat below solar,
while abundances of the $\alpha$-elements are above solar. As an
example, for \element[][]{Fe} we find that about 2/3 of the pixels
show an average abundance $\sim$1/2 solar, and for \element[][]{S}
the average abundance is $\sim$1.7 solar in 40\% of the pixels.
For comparison, Borkowski et al. (1996) derived for the CSM
component abundances  0.4$\times$solar for \element[][]{Fe} and
\element[][]{Ni}, and slightly below solar for other elements
heavier than \element[][]{Ne}.

On the other hand we do see overabundances in particular locations
inside the remnant that lead to global overabundances. From the
results listed in Table~\ref{table2} we can compute for each
element the ratio [X] between the mass fraction per unit solar
mass of element X and the mass fraction of the same element in the
sun. We can compute this overabundance by mass in two different
assumptions: (1) we can consider only pixels in which we detect
the element, or (2) we can consider also the contribution of all
the other pixels where we assume that abundances are solar (i.e.
were not altered by nucleosynthesis in the progenitor star). The
global overabundance factors we find according with these two
assumptions are listed in Table~\ref{table4}.

\begin{table}[h]
\centering \caption{Global overabundance factors by mass. Columns
marked with (1) and (2) refer to different assumptions discussed
in the text.} \label{table4}
\begin{tabular}{lcc}
\noalign{\smallskip} \hline

Element & (1) & (2) \\ \noalign{\smallskip} \hline

~~~[\element[][]{Si}] & 3.3 & 3.1 \\

~~~[\element[][]{S}]  & 5.3 & 5.0 \\

~~~[\element[][]{Ar}] & 6.3 & 4.1 \\

~~~[\element[][]{Ca}] & 6.8 & 4.6 \\

~~~[\element[][]{Fe}] & 1.5 & 1.4 \\

~~~[\element[][]{Ni}] & 7.1 & 3.9 \\

\noalign{\smallskip} \hline
\end{tabular}
\end{table}

As we can see from the right panels of Fig.~\ref{fig5}, all the
elements are particularly overabundant in the south-east. This
region is extending up to the east in \element[][]{Si},
\element[][]{S}, \element[][]{Ar}, and \element[][]{Ca} maps. At
lower level of overabundance these four elements trace a sort of
broken shell, more evident in the case of \element[][]{Ar} and
\element[][]{Ca} maps; similar results were found by Holt et al.
(1994) and Hwang et al. (2000).

\section{Conclusions}
\label{Conclusion}

We have performed the first detailed spatially resolved
spectroscopy of Cas~A in the X-rays, using data taken with the
MECS spectrometer on board the {\it Beppo}SAX Observatory. The
extremely well calibrated PSF in the central region of the MECS
allowed us to perform a spatial deconvolution of our data at full
energy resolution. We eventually generated a data cube consisting
of 182 spectra, each one referring to a squared pixel of
24$\arcsec$ in size, covering a region of $\sim$3$\arcmin$ radius
around the centre of Cas~A.

We performed a fit of the 182 spectra using as models a NEI plasma
and a power law with fixed normalization and index to account for
the high energy tail observed by {\it Beppo}SAX. Statistical
errors on counts have been modified to take into account the
systematic effects of the deconvolution procedure.

In a previous study, reasonable fits of the spatially unresolved
spectrum of Cas~A observed by BeppoSAX were obtained using two
thermal models with different physical and chemical parameters,
indicating the presence of two different components in the
remnant. On the contrary, spatially resolved spectroscopy allowed
us to get statistically significant fits using only one thermal
component per pixel. This means that if two thermal components are
indeed present, they are spatially separated.

We found a rather uniformly distributed $kT$, with a minimum in
the east, a maximum in the west, and most of the values in the range
1--2 keV. No evidence is found for high
$kT$ expected from the interaction of the main shock with the ISM.
No anomalies in the $kT$ distribution are found at the edge of the
remnant. We interpret this as an indication that we are detecting
only matter belonging to the progenitor star.

We have then discussed how to re-heat matter to the observed values,
in particular matter enriched by heavier elements, that is expected
to have already cooled down to $<$0.01 keV at the moment of the SN event.
A possibility is to invoke the standard "reverse shock" model.
In this paper we have also suggested a scenario based on azimuthal
secondary shocks in the supersonic flow generated by the main
shock inside the star. Only a small fraction of the kinetic energy
needs to be converted into thermal energy by azimuthal shocks.
The peculiarity of this scenario is that azimuthal shock can be
formed from hours before to years after the SN event.
Actually they could still be at work now. Observations in the X-rays
at much higher spatial resolution could help in clarifying the issue.

From the discussion of our results on the ionisation parameter
$n_{\rm e}t$ we infer that two distinct components there exist:
the first with $n_{\rm e}$ in the range 1--10 cm$^{-3}$, the
second with $n_{\rm e}$ ten times (or more) higher. The first
component (called $a$) can be associated with matter ejected by
the star before the SN event, while the denser component ($b$),
not spherically symmetric, can be associated with matter ejected
during the explosion. An important consequence of this
interpretation is that the mass ratio $M(a)/M(b)$ is $\leq$1/10,
indicating a progenitor star of moderate mass that lost only a
small fraction of the envelope during its pre-SN life.

We also interpret the distribution of component $b$ across the
remnant as an indication that the explosion was not spherically
symmetric, possibly toroidal.

The distribution of abundances indicates that we are detecting a
CSM component with almost solar composition, and an ejecta
component enriched in heavier elements. Abundances found for
$\alpha$--elements are consistent with the current view that Cas~A
was produced by the explosion of a massive star. But, as already
suggested by Vink et al. (1996), low \element[][]{Fe}
overabundances can be an indication that at the moment of the
explosion the mass-cut was rather high, locking most of the
produced \element[][]{Ni}$^{56}$ into the stellar remnant. The
value of \element[][]{Fe} abundance found in the CSM component
(less than the solar one) is again in agreement with theoretical
predictions.

\begin{acknowledgements}
We thank J.Vink and B.Sacco for a critical reading of the
manuscript and for useful comments. This work was supported by ASI
contract ARS--99--15.
\end{acknowledgements}

\begin{table*}[h]
\scriptsize \centering \caption{Results from the NEI model}
\label{table2}
\begin{tabular}{cccccccccccl}
\noalign{\smallskip} \hline
 & & & & & & & & & & & \\
X & Y & $N_{\rm H}$ & $kT$ & $\log(n_{\rm e}t)$ & \element[][]{Si}
& \element[][]{S} & \element[][]{Ar} & \element[][]{Ca} &
\element[][]{Fe} & \element[][]{Ni} & $\chi^{2}_{red}(dof)$ \\
\noalign{\smallskip} \hline
 & & & & & & & & & & & \\

 191 & 321 &     $-$   &     $-$     &     $-$     &     $-$ & $-$     &     $-$     &     $-$     &     $-$     &     $-$ & 0.85(21) \\
 191 & 324 &     $-$     &     $-$     & 11.5$\pm$5.43 &     $-$     &     $-$     &     $-$     &     $-$     &     $-$     &     $-$     & 1.08(35) \\
 191 & 327 & 1.09:  & 0.75$\pm$0.23 & 10.4$\pm$0.35 & 0.92$\pm$0.69 & 1.86$\pm$1.52 & 1.50:  &     $-$     &     $-$     &     $-$     & 1.05(48) \\
 191 & 330 & 1.23:  & 1.01$\pm$0.37 & 10.2$\pm$0.27 & 1.81$\pm$1.35 & 5.28$\pm$3.60 & 7.82:  &     $-$     &     $-$     &     $-$     & 1.00(48) \\
 191 & 333 & 1.31:  & 1.04$\pm$0.52 & 10.3$\pm$0.32 & 1.89$\pm$1.60 & 8.26$\pm$7.48 & 13.2:  &     $-$     &     $-$     &     $-$     & 1.00(45) \\
 191 & 336 &     $-$     & 0.83$\pm$0.52 & 10.4$\pm$0.47 & 1.26:  & 8.24:  &     $-$     & 5.39:  &     $-$     &     $-$     & 1.19(30) \\
 & & & & & & & & & & & \\
 194 & 315 &     $-$     &     $-$     &     $-$     &     $-$     &     $-$     &     $-$     &     $-$     &     $-$     &     $-$     & 0.82(17) \\
 194 & 318 &     $-$     & 1.61$\pm$1.06 & 11.3$\pm$0.61 & 13.5$\pm$11.9 & 8.11$\pm$5.45 &     $-$     & 15.4:  & 2.28$\pm$2.24 &     $-$     & 1.00(48) \\
 194 & 321 &     $-$     & 0.63$\pm$0.10 & 11.6$\pm$1.33 & 1.30$\pm$1.03 & 3.32$\pm$2.52 &     $-$     &     $-$     & 13.5:  &     $-$     & 1.23(73) \\
 194 & 324 & 1.02$\pm$0.99 & 0.99$\pm$0.27 & 11.8$\pm$0.88 & 3.62$\pm$2.71 & 6.43$\pm$4.12 &     $-$     &     $-$     & 1.75:  & 18.6:  & 1.55(91) \\
 194 & 327 & 1.00$\pm$0.99 & 1.22$\pm$0.44 & 11.6$\pm$0.59 & 4.81$\pm$2.82 & 8.93$\pm$4.22 & 9.50$\pm$6.69 & 3.56:  & 0.89:  & 13.3:  & 1.49(91) \\
 194 & 330 & 1.30:  & 1.28$\pm$0.60 & 11.5$\pm$0.70 & 5.77$\pm$3.64 & 10.7$\pm$5.35 & 17.7$\pm$11.0 & 4.90:  & 0.47:  & 10.8:  & 1.23(89) \\
 194 & 333 &     $-$     & 1.35$\pm$0.71 & 11.5$\pm$0.74 & 7.14$\pm$3.87 & 12.2$\pm$3.06 &     $-$     & 11.7$\pm$6.97 & 0.59:  &     $-$     & 1.06(76) \\
 194 & 336 & 1.22:  & 1.09$\pm$0.45 & 10.5$\pm$0.29 & 1.02$\pm$0.58 & 5.97$\pm$3.03 & 6.05:  & 3.54:  & 2.85:  &     $-$     & 1.16(57) \\
 194 & 339 &     $-$     & 1.17$\pm$0.82 & 11.7$\pm$4.07 &     $-$     &     $-$     &     $-$     &     $-$     &     $-$     &     $-$     & 0.77(32) \\
 & & & & & & & & & & & \\
 197 & 312 &     $-$     & 0.50$\pm$0.41 & 11.2$\pm$3.68 &     $-$     &     $-$     &     $-$     &     $-$     &     $-$     &     $-$     & 0.92(22) \\
 197 & 315 &     $-$     & 0.85$\pm$0.18 & 12.2$\pm$3.73 & 4.33$\pm$3.76 & 6.19$\pm$5.12 &     $-$     & 16.8:  & 1.71:  &     $-$     & 0.83(61) \\
 197 & 318 & 1.01:  & 0.87$\pm$0.11 &     $-$     & 3.44$\pm$1.55 & 6.74$\pm$2.98 &     $-$     & 12.2$\pm$9.01 & 4.20$\pm$1.57 & 14.3:  & 1.12(95) \\
 197 & 321 & 1.02$\pm$0.56 & 1.11$\pm$0.24 & 11.7$\pm$0.50 & 7.04$\pm$5.27 & 13.3$\pm$9.32 & 8.35$\pm$7.72 &     $-$     & 6.54:  & 12.1:  & 1.52(113) \\
 197 & 324 & 1.00$\pm$0.94 & 1.21$\pm$0.28 & 11.5$\pm$0.39 & 6.55$\pm$3.25 & 12.9$\pm$5.76 & 9.37$\pm$5.89 & 12.5$\pm$8.43 & 4.54$\pm$3.26 & 11.5:  & 1.72(114) \\
 197 & 327 & 1.19$\pm$0.66 & 1.27$\pm$0.34 & 11.4$\pm$0.39 & 7.49$\pm$3.42 & 14.7$\pm$5.76 & 14.1$\pm$7.31 & 11.8$\pm$7.77 & 2.64$\pm$1.72 & 18.8$\pm$16.9 & 1.19(114) \\
 197 & 330 & 1.22$\pm$0.70 & 1.17$\pm$0.32 & 11.6$\pm$0.54 & 6.30$\pm$3.40 & 11.9$\pm$5.57 & 14.2$\pm$8.36 & 9.62$\pm$7.58 & 1.61$\pm$1.40 & 13.6:  & 0.91(108) \\
 197 & 333 & 1.20$\pm$0.95 & 1.08$\pm$0.11 &     $-$     & 4.61$\pm$2.33 & 7.68$\pm$3.72 & 8.21$\pm$5.58 &     $-$     & 1.12$\pm$0.75 &     $-$     & 0.66(94) \\
 197 & 336 & 1.15$\pm$1.06 & 1.28$\pm$0.40 & 11.9$\pm$1.24 & 6.22$\pm$6.10 & 8.85$\pm$7.02 & 7.94$\pm$7.56 & 10.5:  & 1.02:  &     $-$     & 0.77(79) \\
 197 & 339 & 1.04:  & 1.88$\pm$1.12 & 11.4$\pm$0.62 & 7.11$\pm$6.30 & 7.56$\pm$4.90 & 5.28:  & 7.67:  & 0.61:  &     $-$     & 0.54(55) \\
 197 & 342 &     $-$     & 1.11$\pm$0.65 & 11.7$\pm$1.41 & 1.83:  & 2.40:  &     $-$     &     $-$     &     $-$     &     $-$     & 0.80(22) \\
 & & & & & & & & & & & \\
 200 & 309 &     $-$     & 0.91:  & 11.2$\pm$2.53 &     $-$     &     $-$     &     $-$     &     $-$     &     $-$     &     $-$     & 1.09(17) \\
 200 & 312 & 1.07:  & 1.30$\pm$0.84 & 11.3$\pm$0.76 & 5.09$\pm$4.15 & 4.73$\pm$2.48 &     $-$     &     $-$     & 0.49:  &     $-$     & 0.93(55) \\
 200 & 315 & 1.20$\pm$0.90 & 0.93$\pm$0.11 &     $-$     & 3.72$\pm$1.87 & 6.60$\pm$3.07 & 6.49$\pm$4.97 &     $-$     & 1.56:  &     $-$     & 0.98(94) \\
 200 & 318 & 1.16$\pm$0.52 & 1.12$\pm$0.22 & 11.8$\pm$0.57 & 7.61$\pm$5.22 & 14.9$\pm$9.63 & 17.4$\pm$12.9 & 8.67$\pm$7.88 & 4.69$\pm$4.67 & 7.23:  & 1.43(116) \\
 200 & 321 & 1.16$\pm$0.00 & 1.31$\pm$0.24 & 11.4$\pm$0.31 & 9.18$\pm$3.83 & 18.8$\pm$7.58 &     $-$     & 18.2$\pm$9.64 & 5.12$\pm$3.33 & 9.89:  & 1.75(127) \\
 200 & 324 & 1.18$\pm$0.57 & 1.18$\pm$0.24 & 11.4$\pm$0.29 & 5.31$\pm$1.64 & 11.9$\pm$3.01 & 10.4$\pm$3.91 & 10.4$\pm$5.00 & 2.43$\pm$1.01 & 3.48:  & 1.49(127) \\
 200 & 327 & 1.21$\pm$0.67 & 0.95$\pm$0.18 & 11.6$\pm$0.38 & 3.73$\pm$1.26 & 9.27$\pm$2.77 & 9.28$\pm$4.28 & 7.04$\pm$4.71 & 1.68$\pm$0.97 &     $-$     & 1.19(121) \\
 200 & 330 & 1.26$\pm$0.73 & 0.89$\pm$0.17 & 11.8$\pm$0.63 & 3.37$\pm$1.37 & 7.65$\pm$2.84 & 7.45$\pm$4.37 & 3.87:  & 1.19$\pm$0.94 &     $-$     & 0.95(116) \\
 200 & 333 & 1.23$\pm$0.72 & 0.99$\pm$0.19 & 11.9$\pm$1.01 & 3.40$\pm$1.66 & 5.98$\pm$2.48 & 5.24$\pm$3.51 & 3.60:  & 0.75:  & 6.81:  & 0.84(110) \\
 200 & 336 & 1.06$\pm$0.76 & 1.28$\pm$0.28 & 11.8$\pm$0.57 & 3.26$\pm$1.52 & 3.94$\pm$1.33 & 2.19:  & 3.00:  & 0.53:  & 6.65:  & 0.78(97) \\
 200 & 339 & 1.15$\pm$1.12 & 1.49$\pm$0.46 & 11.6$\pm$0.50 & 3.20$\pm$1.71 & 2.59$\pm$1.12 &     $-$     &     $-$     & 0.41:  & 4.57:  & 0.73(81) \\
 200 & 342 &     $-$     & 1.53$\pm$1.08 & 11.5$\pm$0.87 & 3.33$\pm$3.15 & 1.85:  &     $-$     & 7.37:  &     $-$     & 12.7:  & 1.01(45) \\
 & & & & & & & & & & & \\
 203 & 309 &     $-$     & 1.52$\pm$1.44 & 10.6$\pm$0.83 & 3.04:  & 2.09$\pm$1.87 &     $-$     &     $-$     &     $-$     &     $-$     & 0.97(37) \\
 203 & 312 & 1.20:  & 1.62$\pm$0.67 & 11.1$\pm$0.36 & 4.42$\pm$2.36 & 4.87$\pm$1.66 & 7.28$\pm$3.70 &     $-$     & 0.36$\pm$0.35 &     $-$     & 0.90(77) \\
 203 & 315 & 1.26$\pm$0.66 & 1.01$\pm$0.08 &     $-$     & 4.48$\pm$1.79 & 8.12$\pm$2.85 & 11.8$\pm$5.55 &     $-$     & 1.37$\pm$0.59 &     $-$     & 1.07(110) \\
 203 & 318 & 1.30$\pm$0.64 & 1.18$\pm$0.21 & 11.7$\pm$0.40 & 5.86$\pm$2.27 & 11.7$\pm$3.93 & 14.3$\pm$5.98 & 6.43$\pm$4.28 & 2.12$\pm$1.30 & 2.98:  & 1.42(125) \\
 203 & 321 & 1.30$\pm$0.61 & 1.46$\pm$0.29 & 11.3$\pm$0.22 & 6.46$\pm$1.72 & 12.0$\pm$2.45 & 11.8$\pm$3.39 & 9.43$\pm$3.94 & 1.67$\pm$0.51 & 4.69$\pm$3.59 & 1.45(128) \\
 203 & 324 & 1.28$\pm$0.64 & 1.40$\pm$0.30 & 11.2$\pm$0.21 & 4.56$\pm$1.13 & 8.51$\pm$1.70 & 6.51$\pm$2.29 & 6.55$\pm$3.23 & 0.76$\pm$0.23 & 4.52$\pm$3.40 & 1.21(125) \\
 203 & 327 & 1.20:  & 1.01$\pm$0.18 & 10.4$\pm$0.11 & 1.43$\pm$0.66 & 3.92$\pm$0.85 & 2.33$\pm$1.11 & 0.46:  & 6.69$\pm$5.29 &     $-$     & 0.91(118) \\
 203 & 330 & 1.20:  & 0.86$\pm$0.16 & 10.6$\pm$0.13 & 1.20$\pm$0.75 & 3.42$\pm$1.00 & 1.99$\pm$0.99 & 0.61:  & 7.50:  &     $-$     & 0.88(110) \\
 203 & 333 & 1.20$\pm$0.54 & 0.86$\pm$0.09 & 10.6$\pm$0.13 & 0.90$\pm$0.19 & 2.58$\pm$0.46 & 1.08$\pm$0.75 & 0.75$\pm$0.69 & 8.20$\pm$7.95 &     $-$     & 1.02(109) \\
 203 & 336 & 1.02$\pm$0.68 & 1.24$\pm$0.22 & 11.5$\pm$0.26 & 2.07$\pm$0.58 & 2.89$\pm$0.70 &     $-$     & 2.25:  & 0.39$\pm$0.26 & 6.54$\pm$4.45 & 0.93(106) \\
 203 & 339 & 1.19$\pm$0.81 & 1.54$\pm$0.29 & 11.4$\pm$0.22 & 2.30$\pm$0.77 & 1.95$\pm$0.50 &     $-$     & 1.21:  & 0.26:  &     $-$     & 0.94(94) \\
 203 & 342 &     $-$     & 1.67$\pm$0.63 & 11.4$\pm$0.38 & 2.61$\pm$1.42 & 1.43$\pm$0.74 &     $-$     & 2.97:  &     $-$     &     $-$     & 1.02(62) \\
 203 & 345 &     $-$     & 1.27$\pm$1.14 & 11.4$\pm$1.23 & 2.19:  &     $-$     &     $-$     &     $-$     &     $-$     &     $-$     & 1.05(23) \\
 & & & & & & & & & & & \\
 206 & 309 &     $-$     & 3.97:  & 10.6$\pm$0.30 & 5.04$\pm$4.40 & 3.56$\pm$1.68 & 5.11$\pm$4.37 &     $-$     &     $-$     & 10.1:  & 0.67(53) \\
 206 & 312 & 1.20$\pm$1.20 & 1.99$\pm$0.78 & 11.1$\pm$0.25 & 3.76$\pm$1.72 & 4.59$\pm$1.56 & 5.97$\pm$3.16 &     $-$     & 0.20:  &     $-$     & 0.85(93) \\
 206 & 315 & 1.30$\pm$1.21 & 2.00$\pm$0.42 & 11.1$\pm$0.15 & 4.81$\pm$1.54 & 7.52$\pm$1.66 & 9.32$\pm$2.77 & 4.35$\pm$2.84 & 0.25$\pm$0.22 &     $-$     & 1.15(112) \\
 206 & 318 &     $-$     & 1.82$\pm$0.30 & 11.2$\pm$0.13 & 5.54$\pm$1.61 & 8.16$\pm$1.53 & 7.56$\pm$2.14 & 5.43$\pm$2.53 & 0.40$\pm$0.19 &     $-$     & 1.29(122) \\
 206 & 321 & 1.22$\pm$0.57 & 1.84$\pm$0.33 & 11.1$\pm$0.14 & 4.06$\pm$0.97 & 6.07$\pm$1.07 & 4.22$\pm$1.59 & 4.26$\pm$2.23 & 0.40$\pm$0.18 &     $-$     & 1.34(125) \\
 206 & 324 & 1.12$\pm$0.61 & 1.49$\pm$0.25 & 11.1$\pm$0.15 & 2.47$\pm$0.58 & 3.89$\pm$0.72 & 1.44$\pm$1.21 & 2.03$\pm$1.87 & 0.25$\pm$0.13 & 3.27$\pm$2.32 & 1.07(123) \\
 206 & 327 & 1.00:  & 1.13$\pm$0.12 & 10.3$\pm$0.10 & 1.30$\pm$0.30 & 2.70$\pm$0.44 & 1.10$\pm$0.68 &     $-$     & 5.07$\pm$4.73 &     $-$     & 1.02(113) \\
 206 & 330 & 1.00:  & 0.95$\pm$0.11 & 10.3$\pm$0.10 & 1.66$\pm$0.53 & 3.28$\pm$0.80 & 1.68$\pm$0.96 & 0.79:  &     $-$     &     $-$     & 0.98(111) \\
 206 & 333 & 1.00:  & 0.91$\pm$0.10 & 10.4$\pm$0.10 & 1.36$\pm$0.26 & 3.45$\pm$0.66 & 1.74$\pm$0.97 & 1.63$\pm$0.89 &     $-$     &     $-$     & 1.02(114) \\
 206 & 336 &     $-$     & 1.05$\pm$0.21 & 10.3$\pm$0.12 & 1.15$\pm$0.81 & 3.22$\pm$1.21 & 0.97$\pm$0.93 & 1.78$\pm$1.08 &     $-$     &     $-$     & 0.95(112) \\
 206 & 339 & 1.22$\pm$0.00 & 1.24$\pm$0.10 & 10.3$\pm$0.14 & 0.86$\pm$0.27 & 2.11$\pm$0.54 &     $-$     & 0.55:  & 5.54:  &     $-$     & 1.02(101) \\
 206 & 342 & 1.19$\pm$0.98 & 1.32$\pm$0.26 & 11.5$\pm$0.30 & 2.14$\pm$0.90 & 1.98$\pm$0.63 &     $-$     &     $-$     &     $-$     &     $-$     & 1.06(75) \\
 206 & 345 & 1.19:  & 0.83$\pm$0.25 & 12.1$\pm$4.25 & 1.48$\pm$1.46 & 0.66:  &     $-$     &     $-$     & 0.69:  &     $-$     & 0.95(38) \\
 & & & & & & & & & & & \\
\noalign{\smallskip} \hline
\end{tabular}
\end{table*}

\begin{table*}[h]
\scriptsize {\bf Table~\ref{table2} - {\em continued}}\\
\begin{tabular}{cccccccccccl}
\noalign{\smallskip} \hline
 & & & & & & & & & & & \\
X & Y & $N_{\rm H}$ & $kT$ & $\log(n_{\rm e}t)$ & \element[][]{Si}
& \element[][]{S} & \element[][]{Ar} & \element[][]{Ca} &
\element[][]{Fe} & \element[][]{Ni} & $\chi^{2}_{red}(dof)$ \\
\noalign{\smallskip} \hline
 & & & & & & & & & & & \\
 209 & 306 &     $-$     & 1.39$\pm$0.70 & 11.2$\pm$1.06 & 1.23:  &     $-$     &     $-$     &     $-$     &     $-$     &     $-$     & 1.00(23) \\
 209 & 309 &     $-$     & 1.56$\pm$0.66 & 10.9$\pm$0.48 & 1.00$\pm$0.59 & 1.24$\pm$0.68 &     $-$     &     $-$     &     $-$     & 9.47:  & 0.75(62) \\
 209 & 312 & 1.29$\pm$1.16 & 1.56$\pm$0.35 & 11.3$\pm$0.23 & 1.57$\pm$0.58 & 2.08$\pm$0.55 & 1.00:  &     $-$     & 0.12:  & 1.34:  & 0.77(93) \\
 209 & 315 & 1.38:  & 1.68$\pm$0.35 & 11.3$\pm$0.19 & 2.63$\pm$0.73 & 3.48$\pm$0.78 & 3.92$\pm$1.75 & 2.29$\pm$2.14 & 0.14:  &     $-$     & 0.94(111) \\
 209 & 318 & 1.30$\pm$1.04 & 1.98$\pm$0.33 & 11.2$\pm$0.12 & 3.21$\pm$0.83 & 3.99$\pm$0.77 & 3.93$\pm$1.53 & 3.93$\pm$2.09 & 0.16:  &     $-$     & 0.92(116) \\
 209 & 321 & 1.05$\pm$0.53 & 1.72$\pm$0.24 & 11.3$\pm$0.12 & 2.15$\pm$0.48 & 2.45$\pm$0.48 & 1.36$\pm$1.10 & 1.40:  & 0.18$\pm$0.15 &     $-$     & 0.81(120) \\
 209 & 324 &     $-$     & 1.57$\pm$0.20 & 11.1$\pm$0.14 & 1.46$\pm$0.31 & 1.70$\pm$0.27 &     $-$     &     $-$     & 0.11:  & 1.54:  & 0.85(119) \\
 209 & 327 &     $-$     & 1.36$\pm$0.16 & 10.5$\pm$0.18 & 0.97$\pm$0.17 & 1.70$\pm$0.29 &     $-$     &     $-$     & 0.50:  &     $-$     & 1.00(115) \\
 209 & 330 &     $-$     & 1.18$\pm$0.10 & 10.1$\pm$0.09 & 2.12$\pm$0.70 & 3.08$\pm$0.70 & 1.56$\pm$0.82 &     $-$     & 17.7:  &     $-$     & 1.11(119) \\
 209 & 333 &     $-$     & 1.39$\pm$0.19 & 10.1$\pm$0.07 & 2.98$\pm$0.72 & 6.03$\pm$1.28 & 4.88$\pm$2.21 & 2.62$\pm$1.94 &     $-$     &     $-$     & 1.11(119) \\
 209 & 336 &     $-$     & 1.35$\pm$0.19 & 10.3$\pm$0.07 & 2.20$\pm$0.48 & 6.28$\pm$1.16 & 5.15$\pm$2.22 & 4.57$\pm$2.18 &     $-$     &     $-$     & 1.08(119) \\
 209 & 339 & 1.20:  & 1.13$\pm$0.21 & 10.3$\pm$0.11 & 1.30$\pm$0.82 & 3.99$\pm$1.53 & 2.28$\pm$1.14 & 2.33$\pm$1.25 &     $-$     &     $-$     & 1.05(112) \\
 209 & 342 & 1.30$\pm$1.07 & 1.05$\pm$0.14 & 10.3$\pm$0.16 & 0.71$\pm$0.27 & 1.94$\pm$0.46 & 0.88$\pm$0.84 &     $-$     & 2.64:  &     $-$     & 1.03(88) \\
 209 & 345 & 1.30$\pm$1.06 & 0.77$\pm$0.16 & 10.4$\pm$0.29 & 0.75$\pm$0.43 & 0.97$\pm$0.55 &     $-$     &     $-$     &     $-$     &     $-$     & 1.08(48) \\
 & & & & & & & & & & & \\
 212 & 306 &     $-$     & 1.51$\pm$0.88 & 10.0$\pm$1.26 &     $-$     &     $-$     &     $-$     &     $-$     &     $-$     &     $-$     & 0.97(30) \\
 212 & 309 &     $-$     & 1.55$\pm$0.34 & 10.2$\pm$0.79 & 0.40:  & 0.63:  &     $-$     &     $-$     &     $-$     &     $-$     & 0.74(68) \\
 212 & 312 &     $-$     & 1.45$\pm$0.26 & 11.3$\pm$0.26 & 0.84$\pm$0.37 & 1.27$\pm$0.38 &     $-$     & 2.94$\pm$2.19 & 0.12:  & 4.91$\pm$3.01 & 0.80(94) \\
 212 & 315 & 1.30$\pm$0.91 & 1.70$\pm$0.34 & 11.3$\pm$0.18 & 1.50$\pm$0.46 & 1.95$\pm$0.53 & 1.83$\pm$1.41 & 2.36$\pm$2.03 & 0.18:  & 2.36$\pm$2.14 & 0.87(113) \\
 212 & 318 & 1.20$\pm$0.64 & 1.96$\pm$0.36 & 11.2$\pm$0.14 & 1.85$\pm$0.49 & 2.08$\pm$0.48 & 2.09$\pm$1.28 & 1.87$\pm$1.75 &     $-$     &     $-$     & 0.79(120) \\
 212 & 321 &     $-$     & 1.78$\pm$0.22 & 11.3$\pm$0.12 & 1.37$\pm$0.33 & 1.20$\pm$0.27 & 0.57:  &     $-$     &     $-$     &     $-$     & 0.62(117) \\
 212 & 324 & 1.00$\pm$0.82 & 1.67$\pm$0.24 & 11.2$\pm$0.14 & 1.19$\pm$0.29 & 1.01$\pm$0.25 &     $-$     &     $-$     & 0.08:  & 1.77$\pm$1.50 & 0.56(112) \\
 212 & 327 & 1.00:  & 1.24$\pm$0.13 & 10.1$\pm$0.17 & 1.06$\pm$0.59 & 1.21$\pm$0.40 &     $-$     &     $-$     & 15.4:  &     $-$     & 0.80(116) \\
 212 & 330 &     $-$     & 1.29$\pm$0.11 & 10.1$\pm$0.12 & 1.44$\pm$0.57 & 2.39$\pm$0.66 & 1.93$\pm$0.87 &     $-$     &     $-$     &     $-$     & 0.95(119) \\
 212 & 333 &     $-$     & 1.68$\pm$0.21 & 10.1$\pm$0.06 & 2.66$\pm$0.56 & 5.99$\pm$1.08 & 7.70$\pm$2.73 & 1.26:  &     $-$     &     $-$     & 1.17(123) \\
 212 & 336 & 1.01:  & 1.77$\pm$0.44 & 10.2$\pm$0.09 & 3.39$\pm$1.13 & 9.46$\pm$2.06 & 13.2$\pm$4.86 & 6.56$\pm$3.72 &     $-$     &     $-$     & 1.47(124) \\
 212 & 339 & 1.20$\pm$1.17 & 1.29$\pm$0.27 & 10.3$\pm$0.10 & 1.83$\pm$0.83 & 6.24$\pm$1.52 & 6.43$\pm$2.45 & 3.99$\pm$1.99 &     $-$     &     $-$     & 1.30(122) \\
 212 & 342 & 1.31:  & 1.02$\pm$0.12 & 10.3$\pm$0.11 & 0.85$\pm$0.25 & 2.83$\pm$0.60 & 2.17$\pm$1.08 &     $-$     & 7.48$\pm$7.26 &     $-$     & 1.03(96) \\
 212 & 345 &     $-$     & 0.74$\pm$0.11 & 10.4$\pm$0.21 & 0.88$\pm$0.41 & 1.37$\pm$0.53 &     $-$     &     $-$     &     $-$     &     $-$     & 1.34(59) \\
 212 & 348 &     $-$     & 0.58$\pm$0.07 & 11.8:  &     $-$     &     $-$     &     $-$     &     $-$     &     $-$     &     $-$     & 0.80(22) \\
 & & & & & & & & & & & \\
 215 & 306 &     $-$     & 1.64$\pm$1.06 & 9.22$\pm$0.69 & 5.26:  &     $-$     &     $-$     &     $-$     &     $-$     &     $-$     & 0.53(27) \\
 215 & 309 & 1.00:  & 1.47$\pm$0.37 & 10.6$\pm$0.35 & 0.23:  & 0.45:  &     $-$     & 3.69$\pm$2.64 &     $-$     &     $-$     & 0.65(66) \\
 215 & 312 & 1.23$\pm$0.87 & 1.65$\pm$0.37 & 10.7$\pm$0.39 & 0.36$\pm$0.23 & 0.94$\pm$0.36 &     $-$     & 4.07$\pm$3.35 &     $-$     &     $-$     & 0.73(90) \\
 215 & 315 & 1.22$\pm$0.52 & 1.77$\pm$0.25 & 10.4$\pm$0.30 & 0.38$\pm$0.18 & 1.05$\pm$0.27 & 0.69:  & 1.86$\pm$1.40 &     $-$     &     $-$     & 0.93(113) \\
 215 & 318 & 1.31:  & 1.96$\pm$0.34 & 11.2$\pm$0.14 & 1.18$\pm$0.34 & 1.27$\pm$0.34 & 0.83:  &     $-$     &     $-$     & 1.42:  & 0.85(121) \\
 215 & 321 &     $-$     & 1.93$\pm$0.28 & 11.3$\pm$0.13 & 1.11$\pm$0.29 & 0.91$\pm$0.26 &     $-$     &     $-$     &     $-$     &     $-$     & 0.79(120) \\
 215 & 324 &     $-$     & 1.72$\pm$0.25 & 11.3$\pm$0.14 & 0.94$\pm$0.25 & 0.75$\pm$0.25 &     $-$     &     $-$     & 0.09:  &     $-$     & 0.66(119) \\
 215 & 327 &     $-$     & 1.28$\pm$0.09 & 10.2$\pm$0.19 & 0.56$\pm$0.41 & 0.88$\pm$0.32 &     $-$     &     $-$     & 12.6:  &     $-$     & 0.75(120) \\
 215 & 330 &     $-$     & 1.17$\pm$0.17 & 11.0$\pm$0.22 & 0.61$\pm$0.13 & 1.39$\pm$0.25 & 1.76$\pm$0.96 &     $-$     & 0.66$\pm$0.32 &     $-$     & 0.92(126) \\
 215 & 333 &     $-$     & 1.48$\pm$0.12 & 10.3$\pm$0.10 & 1.21$\pm$0.22 & 3.41$\pm$0.54 & 4.44$\pm$1.13 &     $-$     & 11.8$\pm$9.10 &     $-$     & 1.15(128) \\
 215 & 336 & 1.00:  & 2.28$\pm$0.45 & 10.4$\pm$0.07 & 3.10$\pm$0.71 & 7.84$\pm$1.15 & 11.8$\pm$2.84 & 7.74$\pm$3.16 & 5.84$\pm$1.96 &     $-$     & 1.34(131) \\
 215 & 339 & 1.20$\pm$0.71 & 1.83$\pm$0.36 & 10.4$\pm$0.09 & 2.36$\pm$0.60 & 6.98$\pm$1.10 & 9.72$\pm$2.65 & 7.07$\pm$3.10 & 4.50$\pm$1.96 &     $-$     & 1.45(125) \\
 215 & 342 & 1.21$\pm$0.64 & 1.25$\pm$0.19 & 10.4$\pm$0.12 & 1.10$\pm$0.24 & 3.80$\pm$0.71 & 3.31$\pm$1.59 & 1.03$\pm$1.00 & 2.42$\pm$2.16 &     $-$     & 1.01(109) \\
 215 & 345 &     $-$     & 0.81$\pm$0.12 & 10.3$\pm$0.18 & 0.92$\pm$0.36 & 1.82$\pm$0.55 &     $-$     &     $-$     &     $-$     &     $-$     & 0.94(68) \\
 215 & 348 &     $-$     & 0.58$\pm$0.38 & 11.5:  &     $-$     &     $-$     &     $-$     &     $-$     &     $-$     &     $-$     & 0.90(23) \\
 & & & & & & & & & & & \\
 218 & 306 &     $-$     & 1.66$\pm$1.42 & 9.38$\pm$0.48 & 5.04:  & 5.27:  &     $-$     &     $-$     &     $-$     &     $-$     & 0.53(25) \\
 218 & 309 & 1.10:  & 1.43$\pm$0.45 & 9.60$\pm$0.33 & 2.04:  & 1.58:  &     $-$     &     $-$     &     $-$     &     $-$     & 0.73(55) \\
 218 & 312 & 1.21$\pm$0.74 & 1.74$\pm$0.33 & 10.2$\pm$0.42 & 0.39$\pm$0.33 & 1.06$\pm$0.55 &     $-$     & 1.15:  &     $-$     &     $-$     & 0.75(89) \\
 218 & 315 & 1.31:  & 1.90$\pm$0.36 & 10.3$\pm$0.34 & 0.38$\pm$0.24 & 1.04$\pm$0.29 & 0.83:  &     $-$     &     $-$     &     $-$     & 0.82(106) \\
 218 & 318 &     $-$     & 2.14$\pm$0.29 & 11.1$\pm$0.13 & 1.05$\pm$0.32 & 1.30$\pm$0.27 & 0.83:  &     $-$     &     $-$     & 1.08:  & 0.79(117) \\
 218 & 321 & 1.19$\pm$0.49 & 2.23$\pm$0.28 & 11.2$\pm$0.11 & 1.22$\pm$0.32 & 1.27$\pm$0.26 & 0.52:  &     $-$     &     $-$     &     $-$     & 0.88(127) \\
 218 & 324 &     $-$     & 1.88$\pm$0.22 & 11.3$\pm$0.12 & 0.86$\pm$0.23 & 0.91$\pm$0.22 &     $-$     &     $-$     & 0.14$\pm$0.14 &     $-$     & 0.92(127) \\
 218 & 327 & 1.00:  & 1.50$\pm$0.19 & 11.2$\pm$0.15 & 0.64$\pm$0.20 & 0.99$\pm$0.22 & 0.79:  &     $-$     & 0.23$\pm$0.12 & 2.34$\pm$1.42 & 0.96(129) \\
 218 & 330 & 1.00$\pm$0.75 & 1.32$\pm$0.16 & 11.2$\pm$0.16 & 0.73$\pm$0.19 & 1.51$\pm$0.27 & 1.92$\pm$0.87 &     $-$     & 0.54$\pm$0.13 & 6.55$\pm$2.85 & 1.02(133) \\
 218 & 333 &     $-$     & 1.55$\pm$0.21 & 11.0$\pm$0.12 & 1.23$\pm$0.23 & 2.86$\pm$0.39 & 3.55$\pm$0.99 &     $-$     & 1.02$\pm$0.12 & 6.55$\pm$2.67 & 1.07(137) \\
 218 & 336 & 1.00:  & 2.25$\pm$0.41 & 10.8$\pm$0.10 & 2.78$\pm$0.60 & 6.08$\pm$0.94 & 6.58$\pm$1.61 & 5.00$\pm$2.03 & 1.82$\pm$0.35 & 6.95$\pm$2.94 & 1.09(137) \\
 218 & 339 & 1.28$\pm$0.58 & 1.88$\pm$0.36 & 10.9$\pm$0.13 & 3.06$\pm$0.77 & 6.38$\pm$1.17 & 6.16$\pm$1.85 & 8.05$\pm$2.77 & 1.27$\pm$0.25 & 3.47$\pm$2.46 & 1.35(131) \\
 218 & 342 & 1.27$\pm$0.80 & 1.67$\pm$0.41 & 11.1$\pm$0.20 & 2.92$\pm$0.91 & 4.96$\pm$1.22 & 2.79$\pm$1.90 & 5.13$\pm$3.15 & 0.27$\pm$0.19 & 3.53$\pm$3.06 & 1.21(106) \\
 218 & 345 & 1.16:  & 1.33$\pm$0.41 & 11.3$\pm$0.38 & 3.20$\pm$1.76 & 3.25$\pm$1.12 &     $-$     &     $-$     &     $-$     &     $-$     & 0.93(62) \\
 218 & 348 &     $-$     & 0.61$\pm$0.36 & 10.6$\pm$1.83 &     $-$     &     $-$     &     $-$     &     $-$     &     $-$     &     $-$     & 0.76(19) \\
 & & & & & & & & & & & \\
 221 & 309 &     $-$     & 1.38$\pm$0.55 & 10.1$\pm$0.58 &     $-$     & 1.65:  & 1.81:  &     $-$     &     $-$     &     $-$     & 0.62(44) \\
 221 & 312 & 1.21$\pm$0.86 & 1.59$\pm$0.36 & 10.4$\pm$0.38 & 0.45$\pm$0.28 & 1.38$\pm$0.53 &     $-$     &     $-$     &     $-$     &     $-$     & 0.80(76) \\
 221 & 315 & 1.23$\pm$0.79 & 1.73$\pm$0.38 & 10.9$\pm$0.29 & 0.60$\pm$0.34 & 1.55$\pm$0.49 & 2.48$\pm$1.54 & 1.01:  & 0.11:  &     $-$     & 0.73(97) \\
 221 & 318 &     $-$     & 2.05$\pm$0.31 & 11.1$\pm$0.13 & 1.09$\pm$0.34 & 1.79$\pm$0.40 & 2.24$\pm$1.13 &     $-$     & 0.15$\pm$0.14 &     $-$     & 0.88(115) \\
 221 & 321 & 1.20$\pm$0.51 & 2.48$\pm$0.40 & 11.1$\pm$0.11 & 1.31$\pm$0.33 & 1.95$\pm$0.36 & 1.58$\pm$0.95 &     $-$     & 0.25$\pm$0.17 &     $-$     & 0.92(132) \\
 221 & 324 & 1.20$\pm$0.00 & 2.20$\pm$0.18 & 11.2$\pm$0.10 & 1.11$\pm$0.25 & 1.62$\pm$0.29 & 0.96$\pm$0.75 & 1.06:  & 0.36$\pm$0.15 &     $-$     & 0.96(138) \\
 221 & 327 & 1.04$\pm$0.43 & 1.90$\pm$0.24 & 11.2$\pm$0.11 & 0.86$\pm$0.21 & 1.54$\pm$0.28 & 0.89$\pm$0.84 & 0.95:  & 0.35$\pm$0.15 & 0.88:  & 1.09(138) \\
 221 & 330 & 1.04$\pm$0.37 & 1.78$\pm$0.21 & 11.2$\pm$0.11 & 0.98$\pm$0.20 & 1.84$\pm$0.26 & 1.29$\pm$0.70 &     $-$     & 0.42$\pm$0.13 & 2.61$\pm$1.25 & 1.05(138) \\
 221 & 333 & 1.02$\pm$0.43 & 1.86$\pm$0.27 & 11.1$\pm$0.11 & 1.53$\pm$0.30 & 2.85$\pm$0.45 & 2.43$\pm$0.99 &     $-$     & 0.59$\pm$0.15 & 3.59$\pm$1.51 & 1.25(136) \\
 221 & 336 & 1.00$\pm$0.60 & 2.27$\pm$0.44 & 11.0$\pm$0.11 & 2.91$\pm$0.64 & 5.11$\pm$0.86 & 5.06$\pm$1.54 & 5.16$\pm$2.10 & 0.93$\pm$0.21 & 4.79$\pm$2.01 & 1.29(136) \\
 221 & 339 & 1.16$\pm$0.65 & 2.02$\pm$0.46 & 11.0$\pm$0.15 & 3.51$\pm$1.00 & 5.40$\pm$1.17 & 4.71$\pm$1.98 & 7.65$\pm$3.08 & 0.83$\pm$0.25 & 3.35$\pm$2.32 & 1.29(120) \\
 221 & 342 & 1.04:  & 1.81$\pm$0.65 & 11.1$\pm$0.26 & 2.96$\pm$1.24 & 3.96$\pm$1.31 &     $-$     & 2.87:  & 0.21:  &     $-$     & 1.41(91) \\
 221 & 345 &     $-$     & 1.92$\pm$1.22 & 10.9$\pm$0.43 & 4.73$\pm$4.05 & 3.81$\pm$2.07 &     $-$     & 7.10:  &     $-$     &     $-$     & 1.01(47) \\
 & & & & & & & & & & & \\
\noalign{\smallskip} \hline
\end{tabular}
\end{table*}

\begin{table*}[h]
\scriptsize {\bf Table~\ref{table2} - {\em continued}}\\
\begin{tabular}{cccccccccccl}
\noalign{\smallskip} \hline
 & & & & & & & & & & & \\
X & Y & $N_{\rm H}$ & $kT$ & $\log(n_{\rm e}t)$ & \element[][]{Si}
& \element[][]{S} & \element[][]{Ar} & \element[][]{Ca} &
\element[][]{Fe} & \element[][]{Ni} & $\chi^{2}_{red}(dof)$ \\
\noalign{\smallskip} \hline
 & & & & & & & & & & & \\
 224 & 309 &     $-$     & 1.49$\pm$0.87 & 11.1$\pm$0.77 & 1.22$\pm$1.10 & 1.85$\pm$1.30 &     $-$     &     $-$     & 0.44:  &     $-$     & 0.37(31) \\
 224 & 312 & 1.27:  & 1.24$\pm$0.45 & 11.1$\pm$0.64 & 0.65$\pm$0.46 & 1.47$\pm$0.74 &     $-$     &     $-$     &     $-$     &     $-$     & 0.61(56) \\
 224 & 315 & 1.31:  & 1.21$\pm$0.19 & 10.1$\pm$0.20 & 0.68$\pm$0.60 & 1.85$\pm$0.85 & 2.68$\pm$1.84 &     $-$     &     $-$     &     $-$     & 0.80(83) \\
 224 & 318 & 1.30$\pm$0.80 & 1.81$\pm$0.32 & 11.2$\pm$0.16 & 1.34$\pm$0.43 & 3.09$\pm$0.70 & 5.55$\pm$1.86 & 4.42$\pm$2.35 & 0.24$\pm$0.18 &     $-$     & 1.19(109) \\
 224 & 321 & 1.20$\pm$0.50 & 3.16$\pm$0.63 & 11.0$\pm$0.10 & 1.99$\pm$0.50 & 4.04$\pm$0.64 & 5.34$\pm$1.41 & 6.12$\pm$1.95 & 0.61$\pm$0.20 &     $-$     & 1.18(135) \\
 224 & 324 & 1.20$\pm$0.41 & 3.07$\pm$0.50 & 11.0$\pm$0.08 & 1.55$\pm$0.34 & 3.15$\pm$0.45 & 3.05$\pm$0.97 & 4.39$\pm$1.46 & 0.73$\pm$0.18 &     $-$     & 1.22(143) \\
 224 & 327 & 1.29$\pm$0.42 & 2.53$\pm$0.33 & 11.0$\pm$0.09 & 1.12$\pm$0.24 & 2.28$\pm$0.32 & 1.78$\pm$0.76 & 3.00$\pm$1.20 & 0.41$\pm$0.14 & 0.89:  & 1.11(143) \\
 224 & 330 & 1.28$\pm$0.42 & 2.11$\pm$0.27 & 11.1$\pm$0.10 & 1.16$\pm$0.24 & 2.04$\pm$0.32 & 1.06$\pm$0.75 & 1.72$\pm$1.16 & 0.32$\pm$0.13 & 2.21$\pm$1.20 & 1.04(138) \\
 224 & 333 & 1.16$\pm$0.50 & 1.77$\pm$0.27 & 11.3$\pm$0.13 & 1.61$\pm$0.36 & 2.51$\pm$0.47 & 1.61$\pm$1.04 & 2.94$\pm$1.61 & 0.37$\pm$0.18 & 3.14$\pm$1.63 & 1.21(129) \\
 224 & 336 & 1.00:  & 1.50$\pm$0.30 & 11.4$\pm$0.23 & 2.25$\pm$0.62 & 3.66$\pm$0.86 & 2.72$\pm$1.73 & 6.22$\pm$2.95 & 0.68$\pm$0.33 & 4.30$\pm$2.85 & 1.16(116) \\
 224 & 339 &     $-$     & 1.39$\pm$0.41 & 11.5$\pm$0.37 & 2.93$\pm$1.17 & 4.27$\pm$1.46 & 2.48:  & 8.71$\pm$5.29 & 0.96$\pm$0.57 & 8.61$\pm$6.85 & 0.88(97) \\
 224 & 342 & 1.14$\pm$1.14 & 1.12$\pm$0.29 & 10.3$\pm$0.30 & 1.00$\pm$0.60 & 2.01$\pm$0.74 &     $-$     & 1.50:  &     $-$     &     $-$     & 0.89(62) \\
 224 & 345 &     $-$     & 1.61$\pm$1.29 & 10.6$\pm$0.75 & 2.13:  & 2.12$\pm$1.83 &     $-$     & 7.12:  &     $-$     &     $-$     & 0.62(24) \\
 & & & & & & & & & & & \\
 227 & 312 &     $-$     & 0.87$\pm$0.38 & 11.2$\pm$1.77 & 0.80:  & 1.30$\pm$1.21 &     $-$     &     $-$     &     $-$     &     $-$     & 0.46(38) \\
 227 & 315 & 1.22$\pm$1.01 & 0.99$\pm$0.21 & 10.8$\pm$0.89 & 0.50$\pm$0.34 & 1.61$\pm$0.72 & 1.29:  &     $-$     &     $-$     &     $-$     & 0.78(64) \\
 227 & 318 & 1.23$\pm$0.88 & 1.07$\pm$0.23 & 11.9$\pm$0.97 & 1.43$\pm$0.73 & 3.78$\pm$1.41 & 7.77$\pm$3.97 & 10.1$\pm$5.78 & 0.47:  & 5.07:  & 1.09(97) \\
 227 & 321 & 1.29$\pm$0.61 & 2.08$\pm$0.45 & 11.2$\pm$0.15 & 1.87$\pm$0.54 & 4.87$\pm$1.05 & 8.57$\pm$2.36 & 10.3$\pm$3.14 & 0.97$\pm$0.29 & 2.76$\pm$2.19 & 1.36(126) \\
 227 & 324 & 1.30$\pm$0.70 & 2.89$\pm$0.47 & 11.0$\pm$0.09 & 1.34$\pm$0.33 & 3.67$\pm$0.55 & 5.12$\pm$1.23 & 6.04$\pm$1.72 & 0.91$\pm$0.20 &     $-$     & 1.38(137) \\
 227 & 327 & 1.30$\pm$0.42 & 2.99$\pm$0.44 & 10.9$\pm$0.09 & 0.93$\pm$0.24 & 2.45$\pm$0.34 & 2.51$\pm$0.82 & 2.86$\pm$1.21 & 0.38$\pm$0.13 &     $-$     & 1.12(137) \\
 227 & 330 & 1.25$\pm$0.51 & 2.28$\pm$0.35 & 11.1$\pm$0.11 & 1.02$\pm$0.27 & 2.01$\pm$0.36 & 1.40$\pm$0.89 & 2.17$\pm$1.38 & 0.13$\pm$0.13 &     $-$     & 0.87(130) \\
 227 & 333 & 1.12$\pm$0.71 & 1.48$\pm$0.27 & 11.5$\pm$0.22 & 1.35$\pm$0.44 & 2.03$\pm$0.56 & 1.55$\pm$1.43 & 3.45$\pm$2.28 & 0.29$\pm$0.25 & 2.52$\pm$2.24 & 0.95(107) \\
 227 & 336 & 1.07$\pm$1.04 & 1.04$\pm$0.23 & 12.1$\pm$1.35 & 1.42$\pm$0.90 & 2.22$\pm$1.07 &     $-$     & 4.23:  & 0.68:  & 5.77:  & 1.06(85) \\
 227 & 339 & 1.17$\pm$1.01 & 0.81$\pm$0.19 & 10.6$\pm$0.29 & 0.57$\pm$0.34 & 1.67$\pm$0.68 &     $-$     & 2.04$\pm$1.73 &     $-$     &     $-$     & 0.78(58) \\
 227 & 342 &     $-$     & 0.68$\pm$0.57 & 11.6$\pm$4.03 &     $-$     &     $-$     &     $-$     &     $-$     &     $-$     &     $-$     & 0.60(28) \\
 & & & & & & & & & & & \\
 230 & 312 &     $-$     & 0.65:  & 10.8$\pm$1.95 &     $-$     &     $-$     &     $-$     &     $-$     &     $-$     &     $-$     & 0.62(16) \\
 230 & 315 & 1.20$\pm$0.00 & 0.71$\pm$0.12 &     $-$     & 1.37$\pm$0.72 & 1.94$\pm$1.10 &     $-$     & 5.61:  &     $-$     &     $-$     & 0.52(40) \\
 230 & 318 & 1.22$\pm$0.92 & 0.94$\pm$0.19 & 10.8$\pm$0.60 & 0.58$\pm$0.28 & 2.23$\pm$0.78 & 1.92:  & 1.26:  &     $-$     &     $-$     & 0.95(69) \\
 230 & 321 &     $-$     & 1.20$\pm$0.16 & 10.2$\pm$0.08 & 0.67$\pm$0.24 & 2.98$\pm$0.69 & 4.01$\pm$1.84 & 2.40$\pm$1.45 &     $-$     &     $-$     & 1.16(102) \\
 230 & 324 & 1.21$\pm$0.45 & 1.88$\pm$0.25 & 10.4$\pm$0.18 & 0.40$\pm$0.16 & 2.12$\pm$0.34 & 4.58$\pm$1.41 & 2.87$\pm$1.55 & 1.20$\pm$1.16 &     $-$     & 1.14(117) \\
 230 & 327 & 1.33:  & 1.81$\pm$0.30 & 11.2$\pm$0.16 & 0.71$\pm$0.30 & 1.91$\pm$0.45 & 3.15$\pm$1.36 & 2.41$\pm$1.76 & 0.28$\pm$0.19 &     $-$     & 0.92(115) \\
 230 & 330 & 1.20$\pm$0.00 & 1.81$\pm$0.25 & 11.2$\pm$0.19 & 0.99$\pm$0.36 & 1.77$\pm$0.44 & 2.82$\pm$1.27 &     $-$     & 0.14:  &     $-$     & 0.87(100) \\
 230 & 333 &     $-$     & 1.46$\pm$0.42 & 11.5$\pm$0.34 & 1.43$\pm$0.76 & 1.61$\pm$0.81 & 1.98:  &     $-$     & 0.30:  &     $-$     & 0.95(73) \\
 230 & 336 &     $-$     & 1.18$\pm$0.60 & 11.7$\pm$0.97 & 1.95$\pm$1.71 & 1.59$\pm$1.40 &     $-$     &     $-$     & 0.66:  &     $-$     & 0.81(44) \\
 230 & 339 &     $-$     & 1.66$\pm$1.56 & 11.4$\pm$0.85 & 5.55:  & 3.87:  &     $-$     &     $-$     & 1.73:  &     $-$     & 0.52(20) \\
 & & & & & & & & & & & \\
 233 & 315 &     $-$     & 0.50$\pm$0.49 & 11.0$\pm$1.80 & 2.20:  &     $-$     &     $-$     &     $-$     &     $-$     &     $-$     & 0.72(18) \\
 233 & 318 & 1.20$\pm$0.00 & 0.55$\pm$0.32 &     $-$     & 1.28:  & 2.78:  &     $-$     &     $-$     &     $-$     &     $-$     & 0.86(40) \\
 233 & 321 &     $-$     & 0.92$\pm$0.25 & 12.2$\pm$2.95 & 1.21$\pm$0.97 & 2.63$\pm$1.31 &     $-$     &     $-$     & 0.68:  &     $-$     & 0.93(63) \\
 233 & 324 & 1.20:  & 1.13$\pm$0.33 & 10.2$\pm$0.21 & 0.56:  & 2.34$\pm$1.87 & 1.32:  &     $-$     &     $-$     &     $-$     & 1.11(78) \\
 233 & 327 &     $-$     & 1.28$\pm$0.37 & 11.4$\pm$0.39 & 0.75$\pm$0.54 & 2.13$\pm$1.02 & 3.26$\pm$3.00 &     $-$     &     $-$     &     $-$     & 0.93(71) \\
 233 & 330 & 1.19:  & 1.27$\pm$0.59 & 11.5$\pm$0.63 & 1.49$\pm$1.12 & 2.36$\pm$1.31 & 3.64:  &     $-$     & 0.27:  &     $-$     & 0.92(57) \\
 233 & 333 &     $-$     & 1.29$\pm$1.09 & 11.4$\pm$0.93 & 1.89:  & 1.83:  &     $-$     &     $-$     &     $-$     &     $-$     & 0.70(32) \\
 & & & & & & & & & & & \\
 236 & 318 &     $-$     &     $-$     &     $-$     &     $-$     &     $-$     &     $-$     &     $-$     &     $-$     &     $-$     & 0.88(13) \\
 236 & 321 &     $-$     &     $-$     &     $-$     &     $-$     &     $-$     &     $-$     &     $-$     &     $-$     &     $-$     & 0.82(24) \\
 236 & 324 &     $-$     & 0.87$\pm$0.33 & 10.4$\pm$0.45 & 0.63:  & 2.11$\pm$1.30 &     $-$     &     $-$     &     $-$     &     $-$     & 0.78(33) \\
 236 & 327 &     $-$     & 0.80$\pm$0.36 & 12.0$\pm$4.44 & 0.92:  & 2.66:  &     $-$     &     $-$     &     $-$     &     $-$     & 0.73(28) \\
 236 & 330 &     $-$     & 1.16$\pm$1.12 & 11.8$\pm$2.10 & 2.71:  & 2.80:  &     $-$     &     $-$     &     $-$     &     $-$     & 0.57(16) \\
 236 & 318 &     $-$     &     $-$     &     $-$     &     $-$     &     $-$     &     $-$     &     $-$     &     $-$     &     $-$     & 0.88(13) \\
 236 & 321 &     $-$     &     $-$     &     $-$     &     $-$     &     $-$     &     $-$     &     $-$     &     $-$     &     $-$     & 0.82(24) \\
 236 & 324 &     $-$     & 0.87$\pm$0.33 & 10.4$\pm$0.45 & 0.63:  & 2.11$\pm$1.30 &     $-$     &     $-$     &     $-$     &     $-$     & 0.78(33) \\
 236 & 327 &     $-$     & 0.80$\pm$0.36 & 12.0$\pm$4.44 & 0.92:  & 2.66:  &     $-$     &     $-$     &     $-$     &     $-$     & 0.73(28) \\
 236 & 330 &     $-$     & 1.16$\pm$1.12 & 11.8$\pm$2.10 & 2.71:  & 2.80:  &     $-$     &     $-$     &     $-$     &     $-$     & 0.57(16) \\
 & & & & & & & & & & & \\
\hline
 & & & & & & & & & & & \\
\multicolumn{12}{l}{\bf Legenda:}\\
 \multicolumn{12}{l}{X and Y are given in sky pixels (unit step 8 arcsec).} \\
 \multicolumn{12}{l}{$N_{\rm H}$ is in units of 10$^{22}$ atoms cm$^{-2}$.} \\
 \multicolumn{12}{l}{$kT$ is in keV.} \\
 \multicolumn{12}{l}{$n_{\rm e}t$ is in units of cm$^{-3}$s.} \\
 \multicolumn{12}{l}{Column marked \element[][]{Si}, \element[][]{S}, \element[][]{Ar},
 \element[][]{Ca}, \element[][]{Fe}, and \element[][]{Ni} list abundance ratios with respect
 to reference values (solar abundances by number).} \\
 \multicolumn{12}{l}{Reference values used in the fitting model are (H=1):} \\
 & & \multicolumn{10}{l}{\element[][]{Si}$=3.77\times10^{-5}$, \element[][]{S}$=1.94\times10^{-5}$,
  \element[][]{Ar}$=3.45\times10^{-6}$, \element[][]{Ca}$=2.3\times10^{-6}$,
  \element[][]{Fe}$=03.38\times10^{-5}$, \element[][]{Ni}$=1.86\times10^{-6}$.} \\
 \multicolumn{12}{l}{Quoted errors are one standard deviation.} \\
 \multicolumn{12}{l}{A value followed by a semicolon indicates a large uncertainty.}\\
 \multicolumn{12}{l}{A dash replaces values that was impossible to constrain to a
 statistically significant level.} \\
 & & & & & & & & & & & \\
 \hline
 \end{tabular}
\end{table*}

\appendix

\section{Error Treatment}
\label{Error-Treatment}

Errors associated to the counts in each pixel of the final images are
the combination of the canonical statistical error (squared root of the
counts) and of a contribution introduced by the deconvolution that depends
on the statistics, on the distribution of counts within
the image, on the PSF approximation,
and on the convergence criteria.
We evaluated the total errors empirically assuming that:
\begin{itemize}
\item[I - ]
deconvolved images relative to energies closer than the instrumental
spectral resolution must be identical within the errors;
\item[II - ]
in each spectrum accumulated from the deconvolved images,
the spread of the points respect to a model that well describe the
overall behaviour (giving a flat distribution of the residuals)
must be within the errors.
\end{itemize}
Each one of these two hypotheses can separately be used to
evaluate the errors, and the results derived from one hypothesis
must automatically satisfy the other one. We have implemented two
procedures, one based on the images (I) and the other on the
spectra (II), and then we have compared the two sets of results.

\subsection{Method I: images based}
\label{Errors-Method-I}

\begin{figure}
\centerline{ \hbox{ \psfig{figure=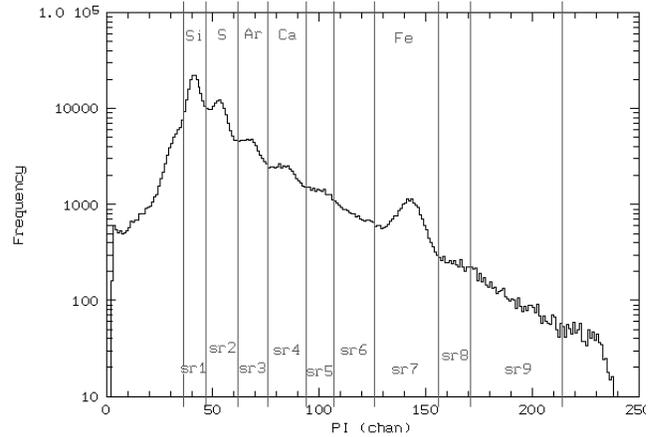,width=8.5cm,clip=} }}
\caption{The spectral regions used to estimate errors superimposed
on data from the ME3 unit, Op2990.} \label{fig6}
\end{figure}

Following the hypothesis that deconvolved images close in energy
must be identical within the errors, we identified 9 energy
ranges, most of which are coincident with the source lines, where
the spatial distribution of the source emission does not vary with
energy. The boundaries chosen for these 9 spectral regions are
shown in Fig.~\ref{fig6}.

Taking into account only the well deconvolved spatial area, we
have produced 9 templates (one for each spectral region) adding
all images in a range and normalizing the source counts in the
summed image. The deconvolved images of each PI in the same
spatial area have then been compared to the relative template
scaled for the source counts in that area by using the variable
$\chi^2_{st\_ima}$ and assigning the statistical errors to the
pixel counts:

$$\chi^2_{st\_ima} = \frac{1}{N_{dof}} \sum_{i,j} \frac{(D_{i,j}-T_{i,j})^2}{D_{i,j}} $$

\noindent
where $D_{i,j}$ refers to the counts in the pixel $(i,j)$ of the
deconvolved PI image, $T_{i,j}$ is the corresponding value in the
related template, and $N_{dof}$ (degrees of freedom) is the number of pixels
on which $\chi^2_{st\_ima}$ is evaluated. Only pixels with
more than 15 counts have been included in the computation.
The variable $\chi^2_{st\_ima}$ is related to the final errors through
a few simple assumptions:
\begin{itemize}
\item
the final errors $Err$ are related to the statistical ones by:

$$Err = \alpha_{ima}~Err_{st}$$

\noindent
where $Err_{st}$ are the statistical errors
and $\alpha_{ima}$ is the correction factor, always greater than 1;
\item
the deconvolution  introduces errors that at first order do not depend
on the position in the image, i.e. $\alpha_{ima}$  is constant over
each PI image.
\end{itemize}
Following these two assumptions, and being the expected value of the
reduced $\chi^2_{red}$ equal to 1, we obtain:

$$\alpha_{ima} = \sqrt \chi^2_{st\_ima} $$

\begin{figure}
\centerline{ \hbox{ \psfig{figure=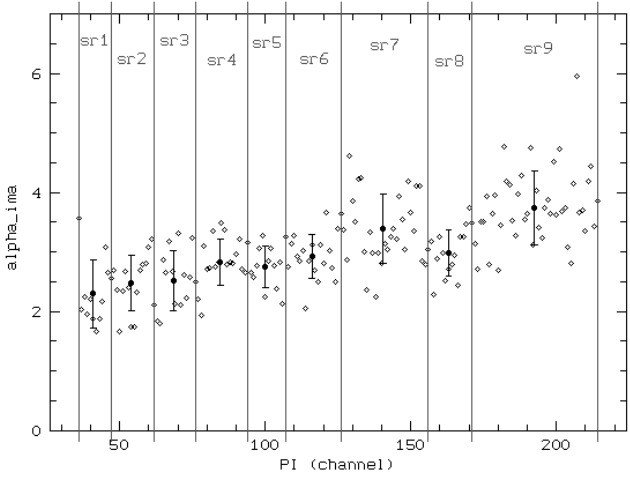,width=8.5cm,clip=} }}
\caption{The $\alpha_{ima}$ correction factor as a function of the
PI channels. Each point refers to a single image. The average
value of $\alpha_{ima}$ in each template region is also shown with
its relative error (root-mean-square deviation).} \label{fig7}
\end{figure}

Fig.~\ref{fig7} plots the $\alpha_{ima}$ values found at the
different PI channels. The average value of $\alpha_{ima}$ in each
template region is also shown with its relative error
(root-mean-square deviation).

\subsection{Method II: spectra based}

We implemented a second procedure following the hypothesis that in
each spectrum the spread of the points respect to a model, that
well describe the general behaviour, must be within the errors. We
chosen to analyze only spectra having a total number of counts
greater than 4$\times$10$^4$: in these spectra in fact the
statistics of most of the energy channels is sufficiently high to
allows us to use the $\chi^2$ test without rebinning. Actually,
some of the high energy channels present very low or zero counts;
however, for homogeneity and to easily automize the procedure, we
did not rebin spectral channel.

We fitted the selected spectra (a total of 24) using a model that,
although it does not correctly describe the emission mechanism of
the source, gives a flat distribution of the residuals for all the
spectra. The fitting model is composed by a power-law plus 9
lines. All parameters are left free to vary but the line widths
that are kept fixed to zero. The MECS response matrix and the
background spectrum are the same used for the spectral analysis
described in Sect.~\ref{Spectral-Analysis}. The spread of the
residuals with respect to the zero level is not uniform over the
whole range: high energy channels present a wider spread  respect
to the low energy ones.

To compute the final errors we need to set some assumptions:
\begin{itemize}
\item
the final errors $Err$ are related to the statistical ones by the
same formula as in Method I (Sect.~\ref{Errors-Method-I}):

$$Err = \alpha_{spec}~Err_{st}$$

\noindent
where $Err_{st}$ are the statistical errors
and $\alpha_{spec}$ is the correction factor;
\item
the value of $\alpha_{spec}$ is a function of the PI channel and does not
depend on the counts detected in it;
\item
$\alpha_{spec}$ can be approximated to a step function of PI being constant
within each given energy range.
\end {itemize}
Note that these assumptions are equivalent to those used in the image
based method (Sect.~\ref{Errors-Method-I})
and are based on the hypothesis that $\alpha_{spec}$ mainly depends on
the statistics of the deconvolved images.

Following these assumptions we divided our spectra in 9 energy
ranges coincident with the ones used in Method I (see
Fig.~\ref{fig6}). Within each of these ranges $\alpha_{spec}$ is
assumed constant and related to $\chi^2_{st\_spec}$, computed
using the statistical errors, by the following formulas:

$$\chi^2_{st\_spec}(sr_k) = \frac {1}{N_{dof(sr_k)}} \sum_{PI}
\frac {(F(PI)-C(PI))^2} {C(PI)} $$

$$\chi^2_{red}(sr_k) = \frac{1}{\alpha^2_{spec}(sr_k)} \chi^2_{st\_spec}(sr_k) $$

By assuming $\chi^2_{red}(sr_k) = 1$, we obtain

$$\alpha_{spec}(sr_k) = \sqrt \chi^2_{st\_spec}(sr_k)$$

\noindent
where $PI$ is ranging within the given $sr_k$ energy range,
$F(PI)$ are the counts relative to the
fitting model in the PI channel, and
$C(PI)$ are the counts detected in the same channel.
Note that $C(PI)$ corresponds to the value of $D_{i,j}$
(used in Method I) at the given PI channel, being $i,j$ the pixel
where the spectrum has been accumulated.

\begin{figure}
\centerline{ \hbox{
\psfig{figure=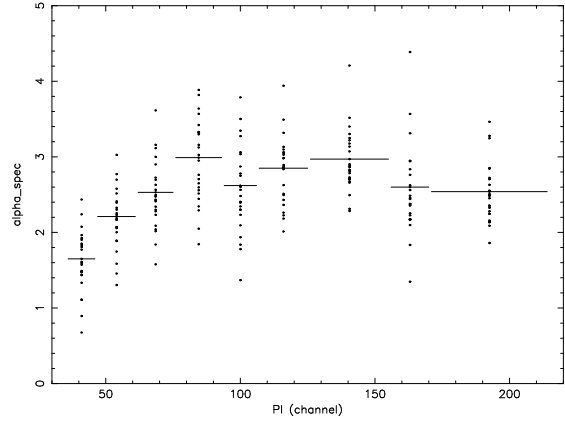,angle=-90,width=8.5cm,clip=} }}
\caption{The $\alpha_{spec}(sr_k)$ correction factor in the 9
defined energy bands. Horizontal bars denote their mean values.
Each point in any $sr_k$ band refers to one of the 24 selected
spectra.} \label{fig8}
\end{figure}

We evaluated the contribution to the $\chi^2_{red}$ in the 9
defined energy bands for each selected spectrum. Following the
assumption that $\alpha_{spec}$ is only function of the PI, we
averaged the 24 values obtained in each of the $sr_k$ regions
assuming as error the root-mean-square deviation from the mean.
Fig.~\ref{fig8} shows the computed $\alpha_{spec}(sr_k)$ together
with their mean value averaged over the 24 analyzed spectra.

\subsection{Correction Factor}

\begin{figure}
\centerline{ \hbox{
\psfig{figure=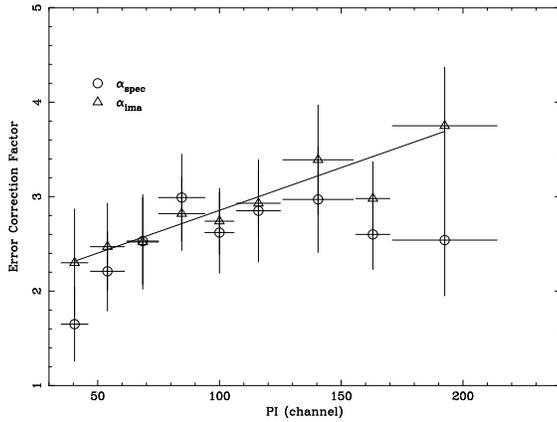,angle=-90,width=8.5cm,clip=} }}
\caption{The $\alpha_{ima}$ and $\alpha_{spec}$ correction factors
at different PI channels. Error bars correspond to the
root-mean-square deviations. The solid line represents the model
used to derive the error correction factor, fitting the
$\alpha_{ima}$ points.} \label{fig9}
\end{figure}

Fig.~\ref{fig9} shows the comparison between the correction
factors $\alpha_{ima}$ and $\alpha_{spec}$ evaluated through the
two methods, image and spectra based, respectively. The values of
$\alpha_{spec}$ and $\alpha_{ima}$ are always compatible within
the errors. The higher discrepancy can be observed in the first
($sr_1$) and last ($sr_9$) point. In the $sr_1$ interval,
$\alpha_{ima}$ is higher than the corresponding $\alpha_{spec}$; a
possible explanation is that this range contains the most
statistical images and the hypothesis that the correction factor
is independent from the counts in the pixel could not be
sufficient: the $\chi^2$ averaged over the whole image is higher
than the one relative to the most statistical pixels. In the case
of the $sr_9$ interval, $\alpha_{spec}$ is lower than the
corresponding $\alpha_{ima}$ being strongly affected by the PI
channels with low or zero counts (these are not included in the
computation of $\alpha_{ima}$).

We then computed the error correction factor from the relation
$\alpha_{ima}$ vs PI channel number (Fig.~\ref{fig7}). Fitting
these points with a line we get the following parameters:

$$\alpha_{ima} = 1.95 + 9.0447\cdot10^{-3} \times PI  $$

The rms of the points respect to the line is 0.5.

\end{document}